\documentclass[12pt]{article}
\usepackage{amsmath,amssymb}
\usepackage{epsfig,euscript,array}
\usepackage{amssymb}
\usepackage{graphics}
\usepackage{graphicx}
\newcommand{\pr}{{\prime}}

\def\beq{\begin{equation}}
\def\eeq{\end{equation}}
\def\beqn{\begin{eqnarray}}
\def\eeqn{\end{eqnarray}}
\def\ba{\begin{eqnarray}}
\def\ea{\end{eqnarray}}

\def\xprim2bar{\overline{x}^{\prime\prime}}

\def\beq{\begin{equation}}
\def\eeq{\end{equation}}

\def\p{\partial}

\newcommand{\beqa}{\begin{eqnarray}}
\newcommand{\eeqa}{\end{eqnarray}}

\newcommand{\be}{\begin{equation}}
\newcommand{\ee}{\end{equation}}
\newcommand{\bea}{\begin{eqnarray}}
\newcommand{\eea}{\end{eqnarray}}

\newcommand{\f}{\frac}

\def\A5{(A_5)_{\rm lat}}
\def\thintablerule{\hrule height0.4pt}
\begin{document}
%%%%%%%%%%%%%%%%%%%%%%%%%%%%
%\rightline{4 March 2011}

\vskip 1.5cm
\centerline{\LARGE Kinetic coupling and dark matter on the lattice}

\vskip 2 cm
\centerline{\large K. Farakos$^{a}$, V.Iconomou$^{a}$ and G. Koutsoumbas$^{a}$}
\vskip1ex
\vskip.5cm
\centerline{\it Department of Physics}
\centerline{\it National Technical University of Athens}
\centerline{\it Zografou Campus, GR-15780 Athens Greece}
\vskip 1.5 true cm
\thintablerule
\vskip 2.0ex
\leftline{\bf Abstract}

We study a $U(1)\otimes U(1)$ system coupled to scalar fields. Initially the model is studied in a novel continuum formulation and study of the appropriate diagonalizations is performed. Three models are examined, in two of which the scalar field couples with both gauge fields while in the third one the scalar field couples only to the dark photon. The model is then treated on a space-time lattice. We determine the phase diagram for various values of the kinetic coupling parameter. Then there follows the determination of masses for the scalar fields and the massive gauge fields, as well as the fine structure constants for the massless gauge fields.

\section{Introduction} \label{s1}

The biggest part of matter in the Universe is of unknown nature.
Our current knowledge suggests that no
more than $5\%$ of the mass-energy of the Universe can be attributed to ordinary matter, while
nearly $27\%$ corresponds to this unknown form of matter known as Dark Matter \cite{c7}. The
remaining $68\%$ correspond to Dark Energy, which is a hypothetical substance driving the
accelerating expansion of the Universe. The nature of dark matter and dark energy are among the greatest
puzzles of modern cosmology and astrophysics. Dark matter has only been observed
through its gravitational effects with ordinary matter and its fundamental
properties are still a mystery. The existence of dark matter suggests the existence of new physics.

Now we present indications about dark matter coming mainly from gravitational observations.
Among the first indications for the existence of dark matter came from the consideration of the
motion of galaxies within the Coma Cluster \cite{c8}. The mass of this cluster has been estimated in two ways,
one based on total luminosity and another based on the virial theorem. The inconsistency of the
two results has given rise to the hypothesis that matter had also some dark component which could not be
detected by the telescopes \cite{c9, c10}.

A rotation curve of stars in galaxies is a measurement of the orbital velocity of stars around the center of the galaxy
as a function of their distance from the galactic center. The rotational velocity was expected to
decrease, however they were measured to be essentially constant at large distances \cite{c1},
so that there should be additional, not visible, i.e. dark, matter, to account for this behaviour.

Gravitational lensing offers extra support to the Dark Matter hypothesis. Study of the images of distant
galaxies may provide clues about the distribution of Dark Matter in space \cite{c11, c12}.
Intra-cluster gas within galaxy clusters may produce X rays, whose lensing results may yield useful
information about the distribution of dark matter \cite{c13, c14}.

The study of the Bullet Cluster \cite{cc12, cc13} provides some of the most compelling observational evidence for the existence of dark matter to date. Further evidence may be found in \cite{cc14}, describing MACS J0025.4-1222 cluster collisions. Evidence in favour of dark matter is also provided by the fluctuations of the Cosmic Microwave Background,
for example, the  Sachs-Wolfe effect \cite{sw}.
Finally, the Universe appears very inhomogeneous today, however CMB was very smooth in the early
Universe. It is difficult to explain the structure formation without the presence of dark matter,
since the magnitude of fluctuations would be too small. Interesting approaches to dark matter in stars have been proposed in \cite{ADM}. For a very partial, but characteristic,
sample of references from the experimental/phenomenological side, on millicharged dark-matter searches, one may consult \cite{exper} and references therein.

Although unrelated to dark matter problem, the idea of a new force under which Standard Model particles are neutral
emerged in 1986 by Holdom \cite{ref1}. Several aspects of this idea have since appeared in the literature in the context of grand unified theories \cite{ref2}, or dark matter models \cite{ref3} or string phenomenology \cite{ref6}. There have been proposals for experimental detection of additional $U(1)$s \cite{ref4}, as well as experimental trials to detect dark photons \cite{ref5}. One may find works on the kinetic coupling and its relation to astrophysics \cite{ref7}. This hidden force has
provided new theoretical and experimental directions in the search for particle dark matter. The extension
of the Standard Model by the addition of the dark $U(1)$ mediated by a dark photon, also known as “hidden
photon”, was originally proposed in \cite{c14}, \cite{c15}. The models can contain a massless dark photon,
with an unbroken $U(1).$
In a theory that includes two $U(1)$ groups, a kinetic mixing term may arise between the two gauge bosons \cite{ref1}. The dark sector is considered as a sector interacting gravitationally, as well as weakly,
with the standard model sector, for example through kinetic mixing of the dark and the visible
gauge field. The gravitational interaction alone is not enough to produce effects which could
serve to detect dark sector particles in the laboratory.

The kinetic mixing between two $U(1)$ gauge groups establishes a
communication channel between the dark and the visible gauge sector. The dark photon acting as
a mediator between the dark and visible sector is known as the “vector portal”, since it is a spin-1
particle. However, other viable portals have been proposed mediated by dark particles of different
spins, such as the “scalar portal” mediated by scalar DM particle or “neutrino portal” mediated by
a dark fermionic field like a right-handed neutrino \cite{c18}.

The dark photon may be chosen to couple to either the electromagnetic or the hyper-charge
of the standard model \cite{italians}, \cite{ref10}. The choice of the appropriate mixing is related to the energy of the
processes under examination: above the electroweak symmetry breaking scale, it is reasonable
to couple to the hypercharge, while below that scale, the coupling to the physical photon is
relevant.

Among the special features of the $U(1)$ models is that the corresponding field strength $F_{\mu\nu}$ is not just covariant, as is the case with the non-abelian models, but it is also gauge invariant. This feature, along with the presence of various $U(1)$ factors, opens up the way to include mixed terms in the Lagrangian, consisting of products of different $U(1)'$s. These considerations arose within the treatment of models of dark matter. If the $U(1)'$s are embedded in a grand unification group, the mixing will be zero at the unification scale, but kinetic coupling may arise through renormalisation group effects at lower scales, where the original group has broken down.

There have been several approaches to the study of models with additional $U(1)$ gauge groups. Kinetic mixing is a recurring theme in these studies. An early one may be found in \cite{ref10}. A nice description of the relevant models may be found in \cite{italians}. This ends up to a new set of decoupled gauge fields. A central issue is the procedure to decouple the two gauge fields, which are coupled through the kinetic coupling, as well as through the symmetry breaking process. It turns out that the transformation is not orthogonal. A by now standard practice is to leave the dark gauge field intact and transform from the visible gauge field to a new one, which contains a small admixture of the dark field. In view of the lattice formulation we propose a different method, which is not orthogonal in the continuum either, however it remains orthogonal in the lattice formalism, the departure from orthogonality been reflected in just a redefinition of lattice parameters. This facilitates the orthogonalization procedure, as will be shown. We describe the two methods of decoupling in the continuum, as well as the decoupling in the lattice formulation of the model. We note that the massive and massless gauge boson cases have been studied separately at tree level, the mass generation mechanism being left unspecified. In this work we employ a higgs field to generate masses for one of the two gauge fields, the other one remaining massless. A unified description of both the massless and the massive gauge bosons follows from this formulation.

This model has been studied by tree level calculations and it is an important issue to proceed to the full quantum model via  non-perturbative techniques, such as lattice Monte Carlo simulations, since symmetry breaking is an essentially non-perturbative effect. To the best of our knowledge, the numerical investigations of such models have not been done in the past, therefore we developed efficient algorithmic software which is applicable on massively parallel computing systems for the Monte Carlo generation of appropriate  configurations and the numerical calculation of the related field-theoretical correlation functions. The point is that, to identify the phases of the resulting model, one should calculate numerically highly non-local quantities, such as the Wilson loops, which will give the potentials felt by heavy quarks in each phase. %Calculating Wilson loops is numerically very demanding, since the lattice used should be very large, if one wants to access the potentials at sufficiently large distances. We used the methodology for the natural parallelization of these models via a natural hypercube splitting of the 4-d lattice and straightforward mapping to the CPU core lattice. Such implementations are well known due to the locality (nearest-neighbor nature) of the action.

This work starts with a novel formulation of the model in the continuum, section \ref{s2}. In particular we present the gauge fields in the presence of the kinetic mixing and present the diagonalization procedure, when matter fields are either missing or present; we identify the massless and massive fields (with their mass) in terms of the original visible and dark fields at the tree level. The results are specified for each of three models (a)model A, where the scalar field charges equal $q_{1s}=1,\ q_{2s} =1,$ (b) model B, where the charges equal $q_{1s}=0,\ q_{2s} =1,$ and (c) model C, with $q_{1s}=q_{2s}=\f{1}{\sqrt{2}}.$ In models A and C both gauge fields are coupled with the scalar field, while in model B the scalar field does not feel the visible gauge field, but only the dark gauge field. Next we present the by now standard formulation of the problem. This procedure is described e.g. in \cite{italians}. This formulation uses a transfomation that  treats the two gauge fields in a quite different way from the beginning, while our procedure is more symmetric with respect to the two fields. We report the results and compare against our formulation. The two procedures yield different linear combinations for the fields after diagonalization in terms of the original fields, as expected, however results, such as the mass of the final massive gauge field turn out to agree completely. Thus the two methods appear to be consistent.

We should note that our model does not permit a distinction of the dark $U(1)$ coupling to the hypercharge or the electromagnetic $U(1).$ To obtain this possibility, we have to use a model based on $SU(2)\times U(1)\times U(1).$ This is deferred to the future. For the time being we examine the two models B and C. The former model B imitates the case where only the dark sector has a spontaneous symmetry breaking, and we study the effects of the symmetry breaking on the remaining unbroken sector. This is rather tricky, in the sense that the diagonalization has to be done numerically. The latter model C imitates the case where both sectors couple to the scalar field with equal strength. Spontaneous symmetry breaking is symmetric and the diagonalization is straightforward.

The formulation on a space-time lattice appears in section \ref{s3}, where we take up the diagonalization procedure in this context. Section \ref{s4} contains  the determination of the higgs phase transition in the lattice formulation for the three models. In particular it contains the Wilson loops and fine structure constants of all models in the symmetric phase. Then the broken phase for model B is described, where the massive and massless gauge fields yield Yukawa mass parameters and fine structure constants respectively. The phase diagram of model C comes next, followed by the plaquettes and correlators of various fields versus $\beta_3,$ as well as their correlators.

Section \ref{s5} contains our conclusions.

\section{The model in the continuum formulation} \label{s2}

The full model in the continuum reads: \be S = \int d^4 x \left[ -\f{1}{4} [{\cal F}_{1\mu\nu} {\cal F}_1^{\mu\nu} + {\cal F}_{2\mu\nu} {\cal F}_2^{\mu\nu} - 2 \kappa {\cal F}_{1\mu\nu} {\cal F}_2^{\mu\nu}]\right. \nonumber \ee
\be \left.  + \left[(\p^\mu - i q_1 g_1 {\cal A}_1^\mu - i q_2 g_2 {\cal A}_2^\mu) \varPhi\right]^\dagger \left[(\p_\mu - i q_1 g_1 {\cal A}_{1 \mu} - i q_2 g_2 {\cal A}_{2 \mu}) \varPhi\right] - \lambda (\varPhi^\dagger \varPhi)^2 + \mu^2 \varPhi^\dagger \varPhi\right]\label{eq1}\ee
where ${\cal F}_{\mu\nu} = \partial_\mu \mathcal A_\nu - \partial_\nu \mathcal A_\mu $ is the Maxwell tensor of $U(1).$

\subsection{Gauge fields}\label{s21}

The gauge sector of the model may be written in the form: %\be {\cal L} = -\f{1}{4} \sum_{0\le \mu<\nu\le 3} {\cal F}_{1 \mu\nu} {\cal F}_{1}^{\mu\nu}-\f{1}{4} \sum_{0\le \mu<\nu\le 3} {\cal F}_{2 \mu\nu} {\cal F}_{2}^{\mu\nu}+\f{\kappa}{2} \sum_{0\le \mu<\nu\le 3} {\cal F}_{1 \mu\nu} {\cal F}_{2}^{\mu\nu},\ee or:
\be {\cal L} = -\f{1}{4} \sum_{0 \le \mu, \nu \le 3} ({\cal F}_{1 \mu\nu},\  {\cal F}_{2 \mu\nu}) \left(\begin{array}{cc} 1 & -\kappa \\ -\kappa & 1 \end{array} \right) \left(\begin{array}{c} {\cal F}_{1}^{\mu\nu}\\  {\cal F}_{2}^{\mu\nu}\end{array} \right).\ee This bilinear form can be made diagonal and yield the alternative expression:
\be {\cal L} =  -\f{1}{4} (1-\kappa) \sum_{0\le \mu,\nu \le 3} {\cal G}_{u, \mu\nu} {\cal G}_{u}^{\mu\nu}-\f{1}{4} (1+\kappa) \sum_{0\le \mu,\nu \le 3} {\cal G}_{v, \mu\nu} {\cal G}_{v}^{\mu\nu},\ee where \be {\cal G}_{u}^{\mu\nu} \equiv \f{1}{\sqrt{2}} ({\cal F}_1^{ \mu\nu}+{\cal F}_2^{\mu\nu}),\  {\cal G}_{v}^{\mu\nu} \equiv \f{1}{\sqrt{2}} (-{\cal F}_1^{ \mu\nu}+{\cal F}_2^{\mu\nu}).\ee One should rescale the fields, to get kinetic terms of the standard form:
\be {\cal A}^\pr_{u \mu} = \sqrt{1-\kappa} {\cal A}_{u \mu}, \ \  {\cal A}^\pr_{v \mu} = \sqrt{1+\kappa} {\cal A}_{v \mu},\ee
\be {\cal G}^\pr_{u, \mu\nu} = \sqrt{1-\kappa} {\cal G}_{u, \mu\nu}, \ \  {\cal G}^\pr_{v, \mu\nu} = \sqrt{1+\kappa} {\cal G}_{v, \mu\nu},\ee so that the Lagrangian density takes the form: \be {\cal L} = -\f{1}{4} \sum_{0\le \mu,\nu \le 3} {\cal G}^\pr_{u, \mu\nu} {\cal G}_{u}^{\pr \mu\nu}-\f{1}{4} \sum_{0\le \mu,\nu \le 3} {\cal G}^\pr_{v, \mu\nu} {\cal G}_{v}^{\pr\mu\nu}.\label{primed}\ee The first step towards diagonalization of the symmetric phase theory involves an orthogonal transformation, however the need to obtain canonical kinetic terms imposes a non-orthogonal factor. Thus the full transformation is the product of the two transformations, that is:
\be {\cal A}^\pr_{u \mu} = \sqrt{\f{1-\kappa}{2}}  ({\cal A}_{1 \mu}+{\cal A}_{2 \mu}),\ \ {\cal A}^\pr_{v \mu} = \sqrt{\f{1+\kappa}{2}}  (-{\cal A}_{1 \mu}+{\cal A}_{2 \mu}),\ee or:
\be \left(\begin{array}{c} {\cal A}_{u}^{\pr \mu}\\  {\cal A}_{v}^{\pr \mu}\end{array} \right) = \left(\begin{array}{cc} \sqrt{\f{1-\kappa}{2}} & \sqrt{\f{1-\kappa}{2}}  \\ -\sqrt{\f{1+\kappa}{2}} & \sqrt{\f{1+\kappa}{2}} \end{array} \right)\left(\begin{array}{c} {\cal A}_{1}^{\mu}\\  {\cal A}_{2}^{\mu}\end{array} \right).\ee It is obvious that the transformation matrix is not orthogonal. We may also write down the inverse relation:
\be \left(\begin{array}{c} {\cal A}_{1}^{\mu}\\  {\cal A}_{2}^{\mu}\end{array} \right) = \left(\begin{array}{cc} \f{1}{\sqrt{2}} \sqrt{\f{1}{1-\kappa}} & -\f{1}{\sqrt{2}} \sqrt{\f{1}{1+\kappa}}  \\ \f{1}{\sqrt{2}} \sqrt{\f{1}{1-\kappa}} & \f{1}{\sqrt{2}} \sqrt{\f{1}{1+\kappa}} \end{array} \right)\left(\begin{array}{c} {\cal A}_{u}^{\pr \mu}\\  {\cal A}_{v}^{\pr \mu}\end{array} \right).\ee

\subsection{Matter fields}\label{s22}

Let us next examine the coupling to the scalar fields, in particular the covariant derivative: \be D_\mu\varPhi = (\p_\mu - i g_1 q_{1s} {\cal A}_{1\mu}  - i g_2 q_{2s} {\cal A}_{2\mu})\varPhi.\ee The dimensionless quantities $q_{1s},\ q_{2 s}$ represent the charge of the scalar fields in units of the respective couplings $g_1,\ g_2.$
The part which will give the mass terms reads:
\be (D_\mu\varPhi)^\dagger (D_\mu\varPhi) \to (-i) (i) \varPhi^\dagger \varPhi (g_1 q_{1s} {\cal A}_{1\mu}  + g_2 q_{2s} {\cal A}_{2\mu}) (g_1 q_{1s} {\cal A}_1^{\mu}  + g_2 q_{2s} {\cal A}_2^{\mu}).\ee The mass of the scalar field derives from the scalar potential \be V(\varPhi^\dagger \varPhi) = \lambda (\varPhi^\dagger \varPhi)^2 - \mu^2 \varPhi^\dagger \varPhi,\ee which has been introduced in equation (\ref{eq1}). Minimization of this expreszion yields the equality: \be v^2=\f{\mu^2}{\lambda}.\ee We may rewrite the above expression in terms of the primed fields and set $<\varPhi^\dagger \varPhi> = \f{v^2}{2},$ so that the mass term will read:
\be \left( {\cal A}_{u \mu}^{\pr},\  {\cal A}_{v \mu}^{\pr} \right) \left(\begin{array}{cc} \f{v^2 \tilde{g}_u^2}{4 (1-\kappa)} & \f{v^2 \tilde{g}_u \tilde{g}_v}{4 \sqrt{1-\kappa^2}}   \\ \f{v^2 \tilde{g}_u \tilde{g}_v}{4 \sqrt{1-\kappa^2}} & \f{v^2 \tilde{g}_v^2}{4 (1+\kappa)} \end{array} \right)\left(\begin{array}{c} {\cal A}_{u}^{\pr \mu}\\  {\cal A}_{v}^{\pr \mu}\end{array} \right),\ \ \tilde{g}_u\equiv g_1 q_{1s} + g_2 q_{2 s},\ \  \tilde{g}_v\equiv -g_1 q_{1s} + g_2 q_{2 s}.\ee
After diagonalization we get:
\be \f{1}{2} \left( {\cal X}_{1 \mu},\  {\cal X}_{2 \mu} \right) \left(\begin{array}{cc} 0 & 0  \\ 0 & m_{\cal A}^2 \end{array} \right)\left(\begin{array}{c} {\cal X}_{1}^{\mu}\\  {\cal X}_{2}^{\mu}\end{array} \right),\ee where: \be m_{\cal A}^2= \f{v^2}{2}\left(\f{\tilde{g}_u^2}{1-\kappa}+\f{\tilde{g}_v^2}{1+\kappa}\right).\ee In the following, in the interest of simplicity, we set $g_1=g_2\equiv g.$ The mass term for the gauge fields reads then: \be \f{1}{2} m_{\cal A}^2 {\cal X}_{2 \mu} {\cal X}_{2}^{\mu}, \ m_{\cal A}^2= \f{g^2 v^2}{2}\left(\f{(q_{1s}+q_{2s})^2}{1-\kappa}+\f{(-q_{1s}+q_{2s})^2}{1+\kappa}\right). \label{gfmasses}\ee
We may express the final potentials, ${\cal X}_1^\mu$ and ${\cal X}_2^\mu,$ in terms of the original potentials, ${\cal A}_1^\mu$ and ${\cal A}_2^\mu,$ via the formulae:

\be \ {\cal X}_1^\mu = \f{1}{\sqrt{q_{1s}^2+q_{2s}^2 + 2 \kappa q_{1s} q_{2s} }} [-(\kappa q_{1s} +q_{2s}) {\cal A}_1^\mu + (q_{1s} +\kappa q_{2s}) {\cal A}_2^\mu],\ee

\be {\cal X}_2^\mu = \f{\sqrt{1-\kappa^2}}{\sqrt{q_{1s}^2+q_{2s}^2 + 2 \kappa q_{1s} q_{2s} }} [q_{1s} {\cal A}_1^\mu + q_{2s} {\cal A}_2^\mu],\ee

The field ${\cal X}_1^\mu$ is massless, while field ${\cal X}_2^\mu$ is massive. We also write down the inverse relations:
\be {\cal A}_1^\mu= \f{-\sqrt{1-\kappa^2} q_{2s} {\cal X}_1^\mu + (q_{1s}+\kappa  q_{2s}) {\cal X}_2^\mu}{\sqrt{1-\kappa^2}\sqrt{q_{1s}^2+q_{2s}^2 + 2 \kappa q_{1s} q_{2s}}},\ee
\be {\cal A}_2^\mu= \f{\sqrt{1-\kappa^2} q_{1s} {\cal X}_1^\mu + (q_{2s}+\kappa  q_{1s}) {\cal X}_2^\mu}{\sqrt{1-\kappa^2}\sqrt{q_{1s}^2+q_{2s}^2 + 2 \kappa q_{1s} q_{2s}}}.\ee

{\bf Model A:}

For the values $g_1=g_2\equiv g, q_{1s}=q_{2s}=1,$ pertaining to model A, we get:
\be  {\cal X}_1^\mu = \sqrt{\f{1 + \kappa}{2}} [-{\cal A}_1^\mu + {\cal A}_2^\mu],\ \ \ {\cal X}_2^\mu = \sqrt{\f{1 - \kappa}{2}} [{\cal A}_1^\mu + {\cal A}_2^\mu],\ee
\be m_{\cal A}^2 = \f{2 g^2 v^2}{1-\kappa}.\label{mA1} \ee
We may also express the ${\cal A}_1^\mu$ and ${\cal A}_2^\mu$ fields in terms of ${\cal X}_1^\mu$ and ${\cal X}_2^\mu:$
\be {\cal A}_1^\mu = -\f{{\cal X}_1^\mu}{\sqrt{2 (1+\kappa)}} + \f{{\cal X}_2^\mu}{\sqrt{2 (1-\kappa)}},\ \ {\cal A}_2^\mu =  \f{{\cal X}_1^\mu}{\sqrt{2 (1+\kappa)}} + \f{{\cal X}_2^\mu}{\sqrt{2 (1-\kappa)}}.\ee
The interaction Lagrangian density for fermions may be written in the form: \be {\cal L}_{A,\psi} = g {\cal J}_\mu^{1 f} {\cal A}_1^\mu + g {\cal J}_\mu^{2 f} {\cal A}_2^\mu = \f{g (-{\cal J}^{1f}_\mu+{\cal J}^{2f}_\mu) {\cal X}_1^\mu}{\sqrt{2  (1+\kappa)}} + \f{{g ({\cal J}^{1f}_\mu+{\cal J}^{2f}_\mu)\cal X}_2^\mu}{\sqrt{2 (1-\kappa)}}.\ee
Of course: \be {\cal J}_\mu^{1 f} = \bar{\Psi}_1 \gamma_\mu \Psi_1,\ \ {\cal J}_\mu^{2 f} = \bar{\Psi}_2 \gamma_\mu \Psi_2.\ee

{\bf Model B:}

For the values $g_1=g_2\equiv g, q_{1s}=0, q_{2s}=1,$ pertaining to model B, we get:
\be {\cal X}_1^\mu = -{\cal A}_1^\mu+\kappa{\cal A}_2^\mu,\ \ \ {\cal X}_2^\mu = \sqrt{1-\kappa^2} {\cal A}_2^\mu.\label{Bx1x2}\ee The inverse relations read:
\be {\cal A}_1^\mu = -{\cal X}_1^\mu + \f{\kappa}{\sqrt{1-\kappa^2}} {\cal X}_2^\mu,\ \ {\cal A}_2^\mu = \f{1}{\sqrt{1-\kappa^2}}{\cal X}_2^\mu.\ee The mass eigenvalue equals: \be m_{\cal A}^2  = \f{g^2 v^2}{1 - \kappa^2}.\label{mB1}\ee
The interaction Lagrangian density for fermions may be written in the form: \be {\cal L}_{B,\psi} = g {\cal J}_\mu^{1 f} {\cal A}_1^\mu + g {\cal J}_\mu^{2 f} {\cal A}_2^\mu = -g {\cal J}_\mu^{1 f} {\cal X}_1^\mu + \f{g}{\sqrt{1-\kappa^2}} [\kappa J_\mu^{1 f}+ J_\mu^{2 f}]{\cal X}_2^\mu.\ee

{\bf Model C:}

For the values $g_1=g_2\equiv g, q_{1s}=q_{2s}=\f{1}{\sqrt{2}},$ pertaining to model C, we get:
\be  {\cal X}_1^\mu = \sqrt{\f{1 + \kappa}{2}} [-{\cal A}_1^\mu + {\cal A}_2^\mu],\ \ \ {\cal X}_2^\mu = \sqrt{\f{1 - \kappa}{2}} [{\cal A}_1^\mu + {\cal A}_2^\mu],\ee
\be m_{\cal A}^2 = \f{g^2 v^2}{1-\kappa}.\label{mC1} \ee
We may also express the ${\cal A}_1^\mu$ and ${\cal A}_2^\mu$ fields in terms of ${\cal X}_1^\mu$ and ${\cal X}_2^\mu:$
\be {\cal A}_1^\mu = -\f{{\cal X}_1^\mu}{\sqrt{2 (1+\kappa)}} + \f{{\cal X}_2^\mu}{\sqrt{2 (1-\kappa)}},\ \ {\cal A}_2^\mu =  \f{{\cal X}_1^\mu}{\sqrt{2 (1+\kappa)}} + \f{{\cal X}_2^\mu}{\sqrt{2 (1-\kappa)}}.\ee
The interaction Lagrangian density for fermions may be written in the form: \be {\cal L}_{A,\psi} = g {\cal J}_\mu^{1 f} {\cal A}_1^\mu + g {\cal J}_\mu^{2 f} {\cal A}_2^\mu = \f{g (-{\cal J}^{1f}_\mu+{\cal J}^{2f}_\mu) {\cal X}_1^\mu}{\sqrt{2  (1+\kappa)}} + \f{{g ({\cal J}^{1f}_\mu+{\cal J}^{2f}_\mu)\cal X}_2^\mu}{\sqrt{2 (1-\kappa)}}.\ee

\subsection{Alternative approach} \label{s23}

The approach followed, e.g. in reference \cite{ref9}, is somewhat different from ours. In the following we examine the gauge sector. In \cite{ref9} the authors start with the fields ${\cal F}_1,\ {\cal F}_2$ and rotate towards $F,\ {\cal F}^\pr$ via a specific matrix. We will derive this matrix from the beginning, under the demand that ${\cal F}_1$ should be just a multiple of ${\cal F}^\pr,$ with no admixture of $F:$
\be  \left(\begin{array}{c} {\cal F}_{1,\mu\nu}\\  {\cal F}_{2, \mu\nu}\end{array} \right) =  \left(\begin{array}{cc} a & 0 \\ c & d \end{array} \right) \left(\begin{array}{c} {\cal F}_{\mu\nu}^\pr \\  {\cal F}_{\mu\nu}\end{array} \right)\Rightarrow \left(\begin{array}{c} {\cal F}_{1,\mu\nu} = a {\cal F}^\pr_{\mu\nu} \\  {\cal F}_{2,\mu\nu} = c {\cal F}_{\mu\nu}^\pr + d {\cal F}_{\mu\nu} \end{array} \right).\ee We substitute it into the expression: \be K= {\cal F}_1^{\mu\nu} {\cal F}_{1,\mu\nu} +  {\cal F}_2^{\mu\nu} {\cal F}_{2,\mu\nu} - 2 \kappa {\cal F}_1^{\mu\nu} {\cal F}_{2,\mu\nu}:\ee
\be K = d^2 {\cal F}^{\mu\nu} {\cal F}_{\mu\nu} + (a^2+c^2 - 2 \kappa a c) {\cal F}^{\pr \mu\nu} {\cal F}_{\mu\nu}^\pr+2 (a d + c d) {\cal F}^{\mu\nu} {\cal F}_{\mu\nu}^\pr.\ee This expression should reduce to \be K = {\cal F}^{\mu\nu} {\cal F}_{\mu\nu} + {\cal F}^{\pr \mu\nu} {\cal F}_{\mu\nu}^\pr,\ee so one gets the relations: \be d^2 =1,\  a^2+c^2 - 2 \kappa a c=1,\ a d + c d=0.\ee One may solve this system (with the additional information that, for $\kappa \to 0,$ $a=d=1).$ The result reads:  \be a = \f{1}{\sqrt{1-\kappa^2}},\ c = \f{\kappa}{\sqrt{1-\kappa^2}},\ d=1,\ee that is: \be  \left(\begin{array}{c} {\cal F}_{1,\mu\nu}\\  {\cal F}_{2, \mu\nu}\end{array} \right)  = \left(\begin{array}{cc} \f{1}{\sqrt{1-\kappa^2}} & 0 \\ \f{\kappa}{\sqrt{1-\kappa^2}} & 1 \end{array} \right) \left(\begin{array}{c} {\cal F}_{\mu\nu}^\pr \\  {\cal F}_{\mu\nu}\end{array} \right).\ee

Considering the gauge part of the action, for the case: \be S_g = \f{1}{2} \sum_{0 \le \mu<\nu\le 3} {\cal F}_1^{\mu\nu} {\cal F}_{1,\mu\nu} + \f{1}{2} \sum_{0 \le \mu<\nu\le 3} {\cal F}_2^{\mu\nu} {\cal F}_{2,\mu\nu} - \kappa \sum_{0 \le \mu<\nu\le 3} {\cal F}_1^{\mu\nu} {\cal F}_{2,\mu\nu} \ee \be \to  \f{1}{2} \sum_{0 \le \mu<\nu\le 3}\left[{\cal F}_1^{\mu\nu} {\cal F}_{1,\mu\nu} + {\cal F}_2^{\mu\nu} {\cal F}_{2,\mu\nu} - 2 \kappa {\cal F}_1^{\mu\nu} {\cal F}_{2,\mu\nu}\right].\ee To summarize, the diagonalization procedure is performed through the change of variables: \be {\cal A}_{1, \mu} = \f{1}{\sqrt{1-\kappa^2}} {\cal A}^\pr_\mu,\ \ {\cal A}_{2, \mu} = \f{\kappa}{\sqrt{1-\kappa^2}} {\cal A}^\pr_\mu + {\cal A}_\mu,\ee with their inverses: \be {\cal A}^\pr_\mu = \sqrt{1-\kappa^2} {\cal A}_{1, \mu},\ \ {\cal A}_\mu = {\cal A}_{2, \mu}-\kappa {\cal A}_{1, \mu}.\ee After substitution one gets: \be S_g = \f{1}{2} \sum_{0 \le \mu<\nu\le 3}\left[{\cal F}^{\mu\nu} {\cal F}_{\mu\nu} + {\cal F}^{\pr \mu\nu} {\cal F}_{\mu\nu}^\pr\right],\ {\cal F}^\pr_{ \mu\nu} \equiv \p_\mu {\cal A}^\pr_\nu - \p_\nu {\cal A}^\pr_\mu ,\ {\cal F}_{\mu\nu} \equiv \p_\mu {\cal A}_\nu - \p_\nu {\cal A}_\mu.\ee Following the symmetry breaking, setting $g_1=g_2=g$ for simplicity, the term \be S_{kin,H} = (D_\mu\varPhi)^\dagger (D_\mu\varPhi), D_\mu\varPhi = (\p_\mu - i g q_{1s} {\cal A}_{1,\mu} - i g q_{2s} {\cal A}_{2 \mu})\varPhi \ee  will contribute and we must diagonalize the total action, given by: \be S=S_g+S_{kin,H}+V(\varPhi^\dagger \varPhi).\ee The covariant derivative may be rewritten in the form: \be D_\mu\varPhi = \left[\p_\mu  - i g q_{1s} \f{1}{\sqrt{1-\kappa^2}} {\cal A}^\pr_\mu - i g q_{2s} \left(\f{\kappa}{\sqrt{1-\kappa^2}} {\cal A}^\pr_\mu + {\cal A}_\mu\right) \right]\varPhi.\ee
Thus the mass term resulting from this form reads:
\be g^2 v^2 ({\cal A}_\mu,\ {\cal A}^\pr_\mu) \left(\begin{array}{cc} q_{2s}^2  & \f{q_{2s} (q_{1s}+\kappa q_{2s} )}{\sqrt{1-\kappa^2}} \\  \f{q_{2s} (q_{1s}+\kappa q_{2s} )}{\sqrt{1-\kappa^2}} & \f{(q_{1s}+\kappa q_{2s} ))^2}{1-\kappa^2}\end{array} \right) \left(\begin{array}{c} {\cal A}_\mu \\  {\cal A}^\pr_\mu \end{array} \right) = g^2 v^2 ({\cal A}_\mu,\ {\cal A}^\pr_\mu) M \left(\begin{array}{c} {\cal A}_\mu \\  {\cal A}^\pr_\mu \end{array} \right)\ee
\be M \equiv \left(\begin{array}{cc} q_{2s}^2 & q_\kappa q_{2s} \\ q_\kappa q_{2s} & q_\kappa^2 \end{array} \right),\ \ q_\kappa \equiv \f{q_{1s} + \kappa q_{2s}}{\sqrt{1-\kappa^2}}.\ee One may diagonalize the matrix $M$ through the transformation: \be  W^T M W =   \left(\begin{array}{cc} 0 & 0 \\ 0 & q_\kappa^2+q_{2s}^2 \end{array} \right),\ \ W = W^T = \f{1}{\sqrt{q^2_{2s}+q_\kappa^2}} \left(\begin{array}{cc} -q_\kappa & q_{2s} \\ q_{2s} & q_\kappa \end{array} \right). \label{WWT}\ee Thus the new fields read: \be \left(\begin{array}{c} {\cal X}_{1,\mu} \\ {\cal X}_{2,\mu}\end{array}\right) = W^T \left(\begin{array}{c} {\cal A}_\mu \\ {\cal A}^\pr_\mu \end{array}\right) = \f{1}{\sqrt{q^2_{2s}+q_\kappa^2}} \left(\begin{array}{cc} -q_\kappa & q_{2s} \\ q_{2s} & q_\kappa \end{array} \right)\left(\begin{array}{c} {\cal A}_\mu \\ {\cal A}^\pr_\mu \end{array}\right), \label{X1X2}\ee so that:
\be {\cal X}_{1,\mu} = \f{1}{\sqrt{q_\kappa^2+q^2_{2s}}} (-q_\kappa {\cal A}_\mu + q_{2s} {\cal A}^\pr_\mu),\  \ {\cal X}_{2,\mu} = \f{1}{\sqrt{q_\kappa^2+q^2_{2s}}} (q_{2s} {\cal A}_\mu+ q_\kappa {\cal A}^\pr_\mu).\label{xx}\ee By inspection of equation (\ref{WWT}) and (\ref{X1X2}) it appears that the ${\cal X}_{1,\mu}$ field is massless, while the ${\cal X}_{2,\mu}$ is massive. The inverse relations of equations (\ref{xx}) read:
\be{\cal A}_{\mu} = \f{1}{\sqrt{q_\kappa^2+q^2_{2s}}} (-q_\kappa {\cal X}_{1,\mu} + q_{2s} {\cal X}_{2,\mu}),\  \ {\cal A}^\pr_{\mu} = \f{1}{\sqrt{q_\kappa^2+q^2_{2s}}} (q_{2s} {\cal X}_{1,\mu}+ q_\kappa {\cal X}_{2,\mu})\ee
The non-zero eigenvalue for the gauge fields equals: \be m^2_{{\cal A}} = g^2 v^2 \left(q_\kappa^2+q^2_{2s} \right) = g^2 v^2 \left[ \left(\f{\kappa q_{2s} + q_{1s}}{\sqrt{1-\kappa^2}}\right)^2+q^2_{2s}\right].\ee
The fermionic lagrangian density reads:
\be {\cal L} = g {\cal J}^{1f}_\mu {\cal A}^{\mu}_1 + g {\cal J}^{2 f}_\mu {\cal A}_{2}^{\mu} = \f{g {\cal J}^{1f}_\mu +\kappa g {\cal J}^{2f}_\mu}{\sqrt{1-\kappa^2}} A^{\pr \mu} + g {\cal J}_2^\mu A^{\mu}\ee
\be = g\f{q_{2s} {\cal J}^{1f}_\mu - q_{1s} {\cal J}^{2f}_\mu}{\sqrt{1-\kappa^2}\sqrt{q_\kappa^2+q_{2s}^2}} {\cal X}_1^\mu + g\f{(q_{1s} + \kappa q_{2s}) {\cal J}^{1f}_\mu + (\kappa q_{1s}+q_{2s}) {\cal J}^{2f}_\mu}{(1-\kappa^2)\sqrt{q_\kappa^2+q_{2s}^2}} {\cal X}_2^\mu\ee

We now continue with the special cases.

{\bf Model A:} \be q_{1s}=1,\ \ q_{2s}=1,\ q_\kappa^2 = \f{1+\kappa}{1-\kappa},\ \ m_{\cal A}^2 =\f{2 g^2 v^2}{1-\kappa}\label{amA1}\ee and:
\be {\cal X}_{1,\mu} = \f{-\sqrt{1+\kappa} {\cal A}_\mu+\sqrt{1-\kappa} {\cal A}^\pr_\mu}{\sqrt{2}} = \sqrt{\f{1+\kappa}{2}} [{\cal A}_{1,\mu} - {\cal A}_{2,\mu}],\ee \be {\cal X}_{2,\mu} = \f{\sqrt{1-\kappa} {\cal A}_\mu+\sqrt{1+\kappa} {\cal A}^\pr_\mu}{\sqrt{2}} = \sqrt{\f{1-\kappa}{2}} [{\cal A}_{1,\mu} + {\cal A}_{2,\mu}]\ee We remark that: \be {\cal X}_{1}^\mu  \propto {\cal A}_v^\mu, \ \ {\cal X}_{2,\mu} \propto {\cal A}_u^\mu,\label{AuAvA}\ee so the appropriate fields ${\cal X}_{1,\mu}$ and ${\cal X}_{2,\mu},$  to deal with after symmetry breaking equal ${\cal A}_v^\mu$ and ${\cal A}_u^\mu.$
The fermionic lagrangian density reads: \be {\cal L}_{A,\psi} = g J_\mu^{1 f} {\cal A}_1^\mu + g J_\mu^{2 f} {\cal A}_2^\mu  = \f{g ( {\cal J}^{1f}_\mu - {\cal J}^{2f}_\mu)}{\sqrt{2 (1+\kappa)}}  {\cal X}_1^\mu + \f{g ( {\cal J}^{1f}_\mu + {\cal J}^{2f}_\mu)}{\sqrt{2 (1-\kappa)}} {\cal X}_2^\mu.\ee

{\bf Model B:} \be q_{1s}=0,\ \ q_{2s}=1,\ q_\kappa^2 = \f{\kappa^2}{1-\kappa^2},\ \ q_\kappa^2+q^2_{2s} = \f{1}{1-\kappa^2},\ m_{\cal A}^2 =\f{g^2 v^2}{1-\kappa^2} \label{amB1} \ee so that:
\be {\cal X}_{1,\mu} = {\cal A}_{1,\mu} - \kappa {\cal A}_{2,\mu},\ {\cal X}_{2,\mu} = \sqrt{1-\kappa^2} {\cal A}_{2,\mu}.\ee
In this case the appropriate fields are approximately ${\cal A}_{1,\mu}$ and ${\cal A}_{2,\mu}.$ The approximation is exact for $\kappa=0.$

The fermionic lagrangian density reads: \be {\cal L}_{B,\psi} = g J_\mu^{1 f} {\cal A}_1^\mu + g J_\mu^{2 f} {\cal A}_2^\mu  = g{\cal J}^{1f}_\mu {\cal X}_1^\mu + \f{g ( \kappa {\cal J}^{1f}_\mu + {\cal J}^{2f}_\mu)}{\sqrt{1-\kappa^2}} {\cal X}_2^\mu. \ee

{\bf Model C:} \be q_{1s}=\f{1}{\sqrt{2}},\ \ q_{2s}=\f{1}{\sqrt{2}},\ q_\kappa^2 = \f{1+\kappa}{2(1-\kappa)},\ \ q_\kappa^2+q^2_{1s} = \f{1}{1-\kappa},\ m_{\cal A}^2 =\f{g^2 v^2}{1-\kappa} \label{amC1}\ee so that:
\be {\cal X}_{1,\mu} = \sqrt{\f{1+\kappa}{2}} [{\cal A}_{1,\mu} - {\cal A}_{2,\mu}],\ {\cal X}_{2,\mu} = \sqrt{\f{1-\kappa}{2}} [{\cal A}_{1,\mu} + {\cal A}_{2,\mu}]\ee
Once more: \be {\cal X}_{1,\mu} \propto {\cal A}_v^\mu, \ \ {\cal X}_{2,\mu} \propto {\cal A}_u^\mu,\label{AuAvC}\ee so the appropriate fields ${\cal X}_{1,\mu}$ and ${\cal X}_{2,\mu},$  to deal with after symmetry breaking equal ${\cal A}_v^\mu$ and ${\cal A}_u^\mu.$

The fermionic lagrangian density reads:
\be {\cal L}_{C,\psi} =  g J_\mu^{1 f} {\cal A}_1^\mu + g J_\mu^{2 f} {\cal A}_2^\mu  = \f{g ( {\cal J}^{1f}_\mu - {\cal J}^{2f}_\mu)}{\sqrt{2 (1+\kappa)}}  {\cal X}_1^\mu + \f{g ( {\cal J}^{1f}_\mu + {\cal J}^{2f}_\mu)}{\sqrt{2 (1-\kappa)}} {\cal X}_2^\mu.\ee

\section{Formulation on a space-time lattice} \label{s3}

The model in the continuum reads: \be S = \left[ -\f{1}{4} \sum_{x, 1\le \mu,\nu\le 4} [{\cal F}_{1\mu\nu} {\cal F}_1^{\mu\nu} + {\cal F}_{2\mu\nu} {\cal F}_2^{\mu\nu} -2 \kappa {\cal F}_{1\mu\nu} {\cal F}_2^{\mu\nu}]\right. \nonumber \ee
\be + \sum_{x, 1\le \mu \le 4} \left[(\p^\mu - i g_1 q_{1s} {\cal A}_1^\mu - i g_2 q_{2s} {\cal A}_2^\mu) \varPhi\right]^\dagger \left[(\p_\mu - i g_1 q_{1s} {\cal A}_{1 \mu} - i g_2 q_{2s} {\cal A}_{2 \mu}) \varPhi\right] \nonumber  \ee \be \left.  - \lambda (\varPhi^\dagger \varPhi)^2 + \mu^2 \varPhi^\dagger \varPhi\right]\ee
where ${\cal F}_{\mu\nu} = \partial_\mu \mathcal A_\nu - \partial_\nu \mathcal A_\mu $ is the field strength tensor of the gauge fields. The non-compact Euclidean lattice action of the 4-dimensional dark matter model reads:
\be S = \frac{1}{2} \beta_1\sum_{x,1\le\mu<\nu\le4} F^{\mu\nu}_1(x) F_1^{\mu\nu}(x)+ \frac{1}{2}  \beta_2\ \sum_{x,1\le\mu<\nu\le4} F^{\mu\nu}_2(x)F_2^{\mu\nu}(x) \nonumber \ee
\be -\beta_3 \sum_{x,1\le\mu<\nu\le4} F_1^{\mu\nu}(x)F_2^{\mu\nu}(x) + \sum_x \Phi^*(x)\Phi(x) - \beta_h \sum_x \left[\sum_{1\le \mu\le 4} \Re(\Phi^*(x) U_{x \hat{\mu}} \Phi(x+a \hat{\mu}))\right]\nonumber \ee
\be + \sum_{x} \beta_R (\Phi^\ast(x)\Phi(x)-1)^2 , \ee
$$F^{\mu\nu}_k(x) \equiv a [A_k^{\mu}(x)+A_k^{\nu}(x+a\hat{\mu}) - A_k^{\mu}(x+a\hat{\nu}) - A_k^{\nu}(x)],\ A_k^{\mu}(x) \equiv g_k {\cal A}_k^{\mu}(x),$$
$$U_{x \mu} \equiv e^{i g_1 q_{1s} a {\cal A}_{1 \mu}(x) + i g_2 q_{2s} a {\cal A}_{2 \mu}(x)} = e^{i a q_{1s} A_{1 \mu}(x) + i a q_{2s} A_{2 \mu}(x)},$$ $$\beta_1=\f{1}{g_1^2},\ \beta_2=\f{1}{g_2^2},\  \beta_3=\f{\kappa}{g_1 g_2}=\kappa \sqrt{\beta_1\beta_2},$$
$$\lambda=\f{4 \beta_R}{\beta_h^2},\ \  a \varPhi= \sqrt{\f{\beta_h}{2}} \Phi.$$
The quantities $q_{1s}$ and $q_{2s}$ are numbers which determine the interaction of the scalar field with the two gauge fields. In this paper we will consider the cases: model A, where $q_{1s} = 1, q_{2s} = 1,$ model B, where $q_{1s} = 0, q_{2s} = 1,$ and model C, where $q_{1s}=\f{1}{\sqrt{2}},\ q_{2s}=\f{1}{\sqrt{2}}.$

\subsection{Mixing} \label{mixing}\label{s31}

The gauge part of the Euclidean action for two kinetically coupled $U(1)$ gauge fields may be transcribed from the continuum to the lattice form as follows: \be S_g = \frac{1}{2} \left[\beta_1 \sum_{x, \mu,\nu} F^{\mu\nu}_1(x) F^{\mu\nu}_1(x) +\beta_2 \sum_{x, \mu,\nu} F^{\mu\nu}_2(x) F_2^{\mu\nu}(x) - 2 \beta_3 \sum_{x, \mu,\nu} F^{\mu\nu}_1(x) F_2^{\mu\nu}(x)\right], \label{contL12} \ee  where: \be \beta_1=\f{1}{g_1^2},\ \beta_2=\f{1}{g_2^2},\  \beta_3=\f{\kappa}{g_1 g_2}=\kappa \sqrt{\beta_1\beta_2}.\ee This expression may be diagonalized and take the form:
\be S_g = \frac{1}{2} \sum_{x, \mu,\nu} \left(\begin{array}{cc} G^{\mu\nu}_u(x), & G^{\mu\nu}_v(x) \end{array}\right) \left(\begin{array}{cc} B_u & 0 \\ 0 & B_v \end{array}\right)\left(\begin{array}{c} G^{\mu\nu}_u(x) \\ G^{\mu\nu}_v(x) \end{array}\right),\ee or:
\be S_g = \frac{1}{2} \left[ B_u \sum_{x, \mu,\nu} G^{\mu\nu}_u(x) G_{u,\mu\nu}(x) + B_v \sum_{x, \mu,\nu} G^{\mu\nu}_v(x) G_{v,\mu\nu}(x)\right].\ee
The eigenvalues read: \be B_u=\frac{1}{2} \left[\beta_1+\beta_2 -\sqrt{(\beta_1-\beta_2)^2+4 \beta_3^2}\right],\ B_v=\frac{1}{2} \left[\beta_1+\beta_2 +\sqrt{(\beta_1-\beta_2)^2+4 \beta_3^2}\right]\ee The orthogonal transformation from the field strengths $F^{\mu\nu}_1(x)$ and $F^{\mu\nu}_2(x)$ to the new ones, $G^{\mu\nu}_u(x)$ and $G^{\mu\nu}_v(x)$ is a very complicated expression. In this work we will concentrate in the case $$\beta_1=\beta_2\equiv \beta_g,$$ where the eigenvalues simplify to: \be B_u = \beta_g-\beta_3,\ \ B_v = \beta_g+\beta_3, \label{BuBv} \ee and the transformation is much simpler: \be G_u^{\mu\nu}=\f{1}{\sqrt{2}} (F_1^{\mu\nu}+F_2^{\mu\nu}),\ \ G_v^{\mu\nu} = \f{1}{\sqrt{2}} (-F_1^{\mu\nu}+F_2^{\mu\nu}),\label{Guv}\ee equivalently: \be F_1^{\mu\nu}=\frac{1}{\sqrt{2}} (G_u^{\mu\nu}-G_v^{\mu\nu}),\ F_2^{\mu\nu}=\frac{1}{\sqrt{2}}(G_u^{\mu\nu}+G_v^{\mu\nu}).\label{F12}\ee Let us remark at this point that equations (\ref{Guv}) for the quantities $G_u^{\mu\nu}$ and $G_v^{\mu\nu}$ may also contain an overall minus sign. Of course, the potentials $A_1^\mu$ and $A_2^\mu$ will transform to $A_u^\mu$ and $A_v^\mu$ in a quite similar way: \be A_u^\mu= \f{1}{\sqrt{2}} (A_1^\mu+A_2^\mu),\ \ A_v^\mu= \f{1}{\sqrt{2}} (-A_1^\mu+A_2^\mu),\ee equivalently: \be A_1^\mu=\frac{1}{\sqrt{2}} (A_u^\mu-A_v^\mu),\ A_2^\mu=\frac{1}{\sqrt{2}}(A_u^\mu+A_v^\mu).\label{A12}\ee We also note the definitions: \be G^{\mu\nu}_{u,v} \equiv \p^\mu A^\nu_{u,v} - \p^\nu A^\mu_{u,v}.\ee The new fields are massless just like the old ones.

We now consider interaction with a scalar field, the covariant derivative reading: \be D^\mu \varPhi = [\p^\mu - i q_{1s} A_1^\mu - i q_{2s} A_2^\mu]\varPhi \ee \be = \left[\p^\mu - i q_{1s} \frac{1}{\sqrt{2}} (A_u^\mu-A_v^\mu) - i q_{2s} \frac{1}{\sqrt{2}} (A_u^\mu+A_v^\mu)\right]\varPhi.\ee Notice that the initial expressions e.g. $i g_k q_{ks} {\cal A}_k^\mu,$ have been replaced by $i q_{ks} A_k^\mu,$ since we use the lattice version for the potentials.
The quantities $q_{1s}$ and $q_{2s}$ are pure numbers which determine the strength of the interaction of  the gauge potentials $A_1$ and $A_2$ with the scalar field. We write more compactly:
\be D^\mu \varPhi = \left[\p^\mu - i Q_{us} A_u^\mu - i Q_{vs} A_v^\mu\right]\varPhi,\ee \be Q_{us}\equiv \frac{1}{\sqrt{2}}(q_{1s}+ q_{2s}),\ Q_{vs}\equiv \frac{1}{\sqrt{2}} (-q_{1s}+ q_{2s}).\label{Quv}\ee Thus we expect that, in the symmetric phase, we will have two massless fields, with the new charges reading $Q_{us}$ and $Q_{vs}.$

We recall that in model C we set $q_{1s}=q_{2s}=\f{1}{\sqrt{2}},$ so that $Q_{us}=1$ and $Q_{vs}=0.$ As we remarked previously, see (\ref{AuAvA}) and (\ref{AuAvC}), \be {\cal X}_1^\mu\propto {\cal A}_v^\mu,\ \ {\cal X}_2^\mu\propto {\cal A}_u^\mu.\ee
We expect that the appropriate fields to consider is ${\cal A}_u^\mu,$ which, in the broken phase, will develop a mass (since it is proportional to ${\cal X}_2^\mu,$ which is massive) and $ {\cal A}_v^\mu,$ which will remain massless, since it is a multiple of the massless field ${\cal X}_1^\mu.$

In model B we set $q_{1s}=0,\ q_{2s}=1,$ so that $Q_{us}=\f{1}{\sqrt{2}}$ and $Q_{vs}=\f{1}{\sqrt{2}},$ as is evident from equations (\ref{Quv}). In this case, the appropriate basis to consider for $\beta_3=0,$ is the one consisting of the massive $ {\cal A}_2^\mu$ and the massless $ {\cal A}_1^\mu.$ However, for $\beta_3>0,$ the appropriate basis is a combination of $ {\cal A}_1^\mu$ and  $ {\cal A}_2^\mu,$ depending on the value of $\beta_3.$

\section{Results} \label{s4}

\subsection{Higgs phase transition} \label{s41}

We examine the phase transition between the symmetric and the broken phase for $\beta_3=0.$ We also set: $\beta_1=\beta_2\equiv \beta_g=2,\ \beta_R=0.001.$
In this work we will examine the three different models mentioned above, differing in the values of $q_{1s}$ and $q_{2s}.$
We rewrite the lattice action substituting: \be U_{x \mu}  = e^{i a q_{1s} A_{1 \mu}(x) + i a q_{2s} A_{2 \mu}(x)} = e^{i \theta_{\mu}(x)},\ \ \Phi(x) = \rho(x) e^{i \phi(x)}:\ee
\be S = \frac{1}{2} \beta_1\sum_{x,1\le\mu,\nu\le4} F^{\mu\nu}_1(x) F_1^{\mu\nu}(x)+ \frac{1}{2}  \beta_2\ \sum_{x,1\le\mu,\nu\le4} F^{\mu\nu}_2(x)F_2^{\mu\nu}(x)-\beta_3 \sum_{x,1\le\mu,\nu\le4} F_1^{\mu\nu}(x)F_2^{\mu\nu}(x) \nonumber\ee
\be  - \beta_h \sum_x \left[\sum_{1\le \mu\le 4} \rho(x) \rho(x+\hat{\mu}) \cos[\theta_{\mu}(x) + \phi(x+\hat{\mu}) - \phi(x)] \right] \nonumber \ee
\be + \sum_x \rho^2(x) + \sum_{x} \beta_R (\rho^2(x)-1)^2. \ee
Notice that: \be D^\mu \varPhi = [\p^\mu - i q_{1s} A_1^\mu - i q_{2s} A_2^\mu]\varPhi = \left[\p^\mu - i Q_{us} A_u^\mu - i Q_{vs} A_v^\mu\right]\varPhi,\ee \be Q_{us}\equiv \frac{1}{\sqrt{2}}(q_{1s}+ q_{2s}),\ Q_{vs}\equiv \frac{1}{\sqrt{2}} (-q_{1s}+ q_{2s}),\ee so that: \be U_{x \mu} = e^{i \theta_{\mu}(x)}  = e^{i Q_{us}  A_{u \mu}(x) + i Q_{vs} A_{v \mu}(x)}.\ee
Model A is defined by $q_{1s}=q_{2s}=1.$ In this case $Q_{us} = \sqrt{2},\  Q_{vs} = 0,\ $ $\theta_{\mu}(x)  = \sqrt{2}  A_{u \mu}(x),$ so that the scalar field interacts only with the $A_{u \mu}(x)$ field and it is also convenient to express the gauge kinetic terms as functions of $A_u^\mu$ and $A_v^\mu$:
\be S = \frac{1}{2} \left[ B_u \sum_{\mu,\nu} G^{\mu\nu}_u(x) G_{u,\mu\nu}(x) + B_v \sum_{\mu,\nu} G^{\mu\nu}_v(x) G_{v,\mu\nu}(x)\right]\nonumber \ee
\be - \beta_h \sum_x \left[\sum_{1\le \mu\le 4} \rho(x) \rho(x+\hat{\mu}) \cos[\sqrt{2}  A_{u \mu}(x) + \phi(x+\hat{\mu}) - \phi(x)] \right]\nonumber \ee
\be + \sum_x \rho^2(x) + \sum_{x} \beta_R (\rho^2(x)-1)^2. \ee One may now perform a rescaling of the gauge fields: $A_{u \mu}(x) \to \f{1}{\sqrt{2}} A_{u \mu}(x),\ A_{v \mu}(x) \to \f{1}{\sqrt{2}} A_{v \mu}(x)$ and end up with:
\be S = \frac{1}{2} \left[ \f{B_u}{2} \sum_{\mu,\nu} G^{\mu\nu}_u(x) G_{u,\mu\nu}(x) + \f{B_v}{2} \sum_{\mu,\nu} G^{\mu\nu}_v(x) G_{v,\mu\nu}(x)\right]\nonumber\ee
\be - \beta_h \sum_x \left[\sum_{1\le \mu\le 4} \rho(x) \rho(x+\hat{\mu}) \cos[A_{u \mu}(x) + \phi(x+\hat{\mu}) - \phi(x)] \right] \nonumber\ee
\be + \sum_x \rho^2(x) + \sum_{x} \beta_R [\rho^2(x)-1]^2. \ee
It is apparent that this action describes a non-interacting gauge field $A_{v \mu}(x)$ along with a field $A_{u \mu}(x)$ coupled to the scalar field. The lattice gauge couplings $\f{B_u}{2}=\f{\beta_g}{2}-\f{\beta_3}{2}$ and $\f{B_v}{2}=\f{\beta_g}{2}+\f{\beta_3}{2}$ are $\f{1}{2}$ times the nominal ones. Thus we expect that the phase transition lines in the $\beta_g - \beta_h$ plane will differ significantly from the cases B and C, where we will see that no rescaling is necessary.

We examine model B, defined by $q_{1s}=0,\ q_{2s}=1,$ for which:
\be S = \frac{1}{2} \beta_g \sum_{,x,1\le\mu,\nu\le4} F^{\mu\nu}_1(x) F_1^{\mu\nu}(x)+ \frac{1}{2}  \beta_g \ \sum_{x,1\le\mu,\nu\le4} F^{\mu\nu}_2(x)F_2^{\mu\nu}(x)-\beta_3 \sum_{x,1\le\mu,\nu\le4} F_1^{\mu\nu}(x)F_2^{\mu\nu}(x) \nonumber\ee
\be - \beta_h \sum_x \left[\sum_{1\le \mu\le 4} \rho(x) \rho(x+\hat{\mu}) \cos[A_{2 \mu}(x) + \phi(x+\hat{\mu}) - \phi(x)] \right] \nonumber\ee
\be + \sum_x \rho^2(x) + \sum_{x} \beta_R (\rho^2(x)-1)^2.\ee
This model describes a non-interacting gauge field $A_{1 \mu}(x)$ along with a field $A_{2 \mu}(x)$ coupled to the scalar field. %Since only $A_{2 \mu}(x)$ couples to the scalar field, it is convenient to keep the original gauge fields $A_{1 \mu}(x)$ and $A_{2 \mu}(x)$ in our study of model B.

Model C is defined by $q_{1s}=q_{2s}=\f{1}{\sqrt{2}}.$ In this case $Q_{us} = 1,\  Q_{vs} = 0,\ $ $\theta_{\mu}(x)  = A_{u \mu}(x),$ and no rescaling is needed:
\be S = \frac{1}{2} \left[ B_u \sum_{\mu,\nu} G^{\mu\nu}_u(x) G_{u,\mu\nu}(x) + B_v \sum_{\mu,\nu} G^{\mu\nu}_v(x) G_{v,\mu\nu}(x)\right] \nonumber\ee
\be - \beta_h \sum_x \left[\sum_{1\le \mu\le 4} \rho(x) \rho(x+\hat{\mu}) \cos[A_{u \mu}(x) + \phi(x+\hat{\mu}) - \phi(x)] \right] \nonumber\ee
\be + \sum_x \rho^2(x) + \sum_{x} \beta_R (\rho^2(x)-1)^2.\ee
It is obvious that model C shares some features with model B; the scalar field couples only with field $A_{2 \mu}(x)$ (in model B) and with field $A_{u \mu}(x)$ in model C. Despite the difference in the choice of gauge fields, the mixing coupling $\beta_3$ stays smaller than $\beta_g,$ as it will be observed in the following, for instance in figure \ref{phd_ABC060} below. As $\beta_3$ towards $\beta_g,$ various quantities get increasingly large values.

In the numerical study we examine the quantities \be <F_1^2>\equiv \f{1}{6 N^4}\sum_{x}\sum_{\mu<\nu=1}^4 <F_1^{\mu\nu}(x) F_1^{\mu\nu}(x)>,\ee \be <F_2^2>\equiv \f{1}{6 N^4}\sum_{x}\sum_{\mu<\nu=1}^4 <F_2^{\mu\nu}(x) F_2^{\mu\nu}(x)>.\ee
We also define a new variable considering the part of the action referring to the coupling of the scalar to the gauge fields: \be - \beta_h \sum_x \left[\sum_{1\le \mu\le 4} \Re(\Phi^*(x) U_{x \hat{\mu}} \Phi(x+\hat{\mu}))\right] \ee \be = - \beta_h \sum_x \left(\sum_{1\le \mu\le 4} \rho(x) \rho(x+\hat{\mu}) \cos[\theta_{\mu}(x) + \phi(x+\hat{\mu}) - \phi(x)]\right),\ee where we have set:
\be U_{x \hat{\mu}} = e^{i a q_1 A_{1 \mu}(x) + i a q_2 A_{2 \mu}(x)} = e^{i \theta_{\mu}(x)},\ \Phi(x) = \rho(x) e^{i \varphi(x)}.\ee The definition of the new, angular, variable is: \be\cos({\rm link}) \equiv \f{1}{4 N^4}\sum_{x}\sum_{\mu=1}^4 <\cos[\theta_{\mu}(x) + \phi(x+\hat{\mu}) - \phi(x)]>.\ee

\begin{figure}[ht]
\begin{center}
\includegraphics[scale=0.30,angle=0]{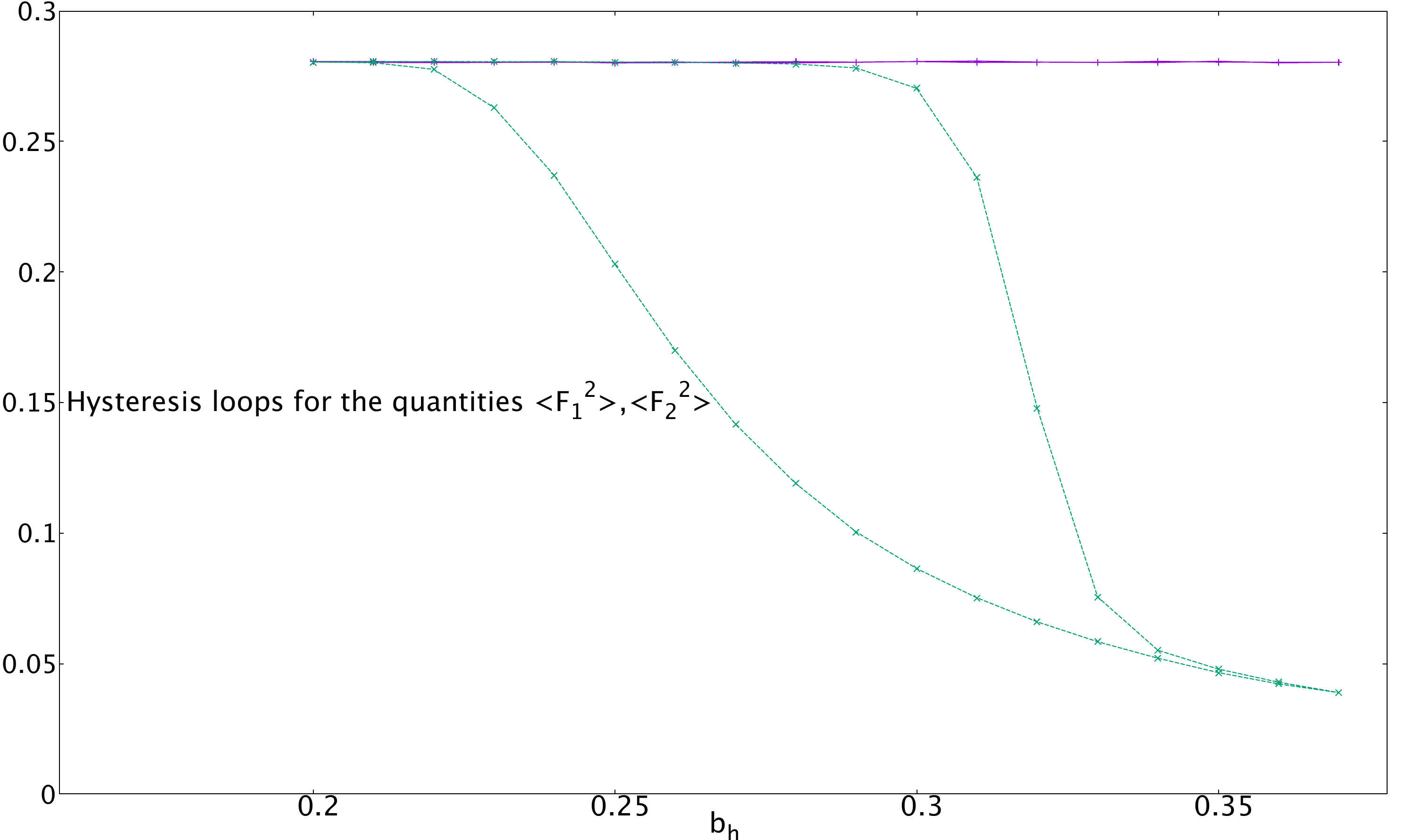}
\end{center}
\caption {Hysteresis loops for the quantities $<F_1^2>$ and $<F_2^2>$ in model B $(b_h\equiv \beta_h).$ The quantity $<F_1^2>$ has no hysteresis loop since, in model B, the $F_1$ field does not couple with the scalar field, so it does not feel the symmetry breaking. The quantity $<F_2^2>$ indicates a first order phase transition.} \label{hyst1}
\end{figure}

\begin{figure}[ht]
\begin{center}
\includegraphics[scale=0.30,angle=0]{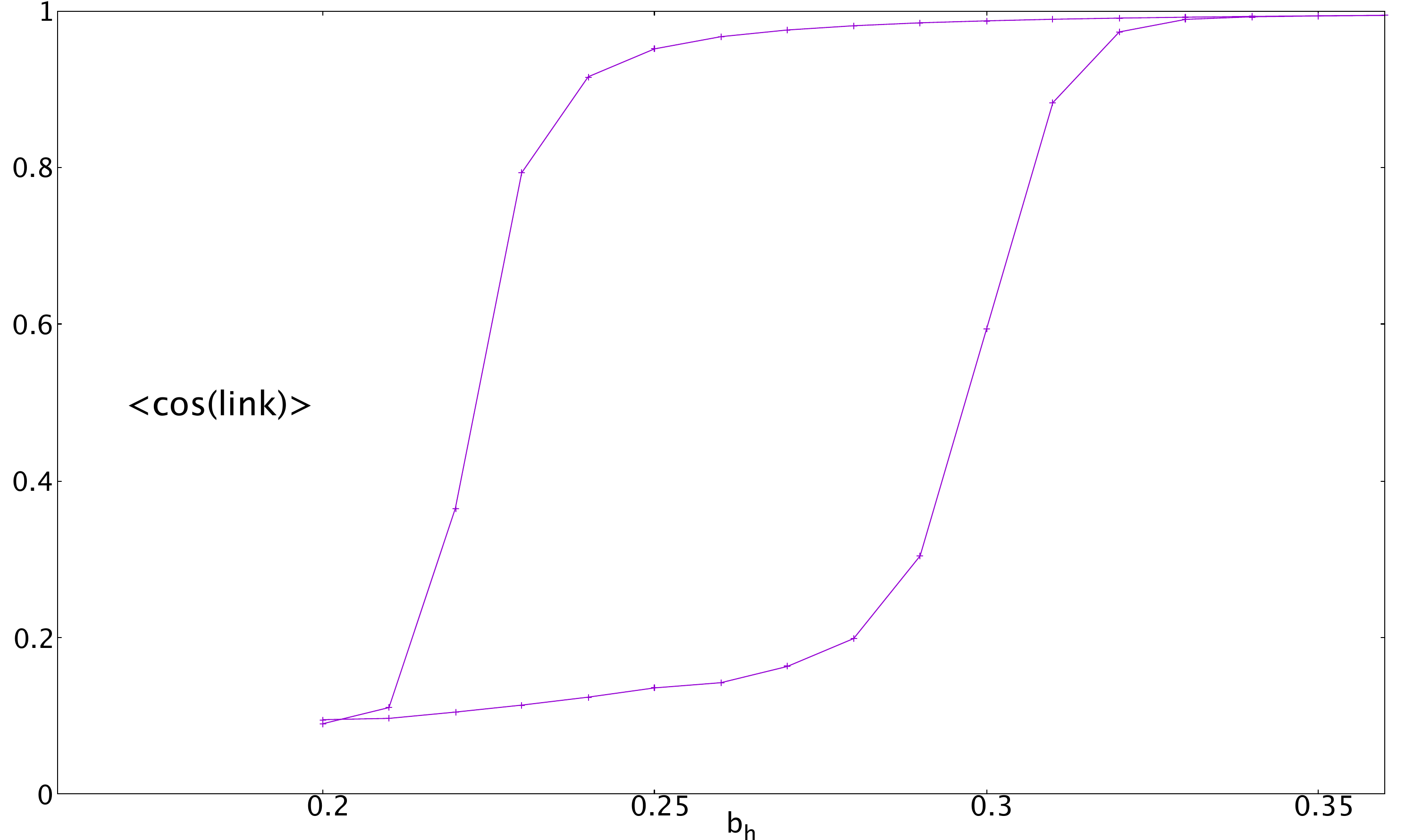}
\end{center}
\caption {Hysteresis loops for ${\rm cos(link)}$ in model B.} \label{hyst2}
\end{figure}

Initially we performed hysteresis loops for constant $\beta_g$ and $\beta_R$ varying $\beta_h$ to locate the phase transition points. One such loop is depicted in figure \ref{hyst1}, which involves the loop for the gauge field in model B and, in figure \ref{hyst2}, the loop for the angle of the scalar field in model B. In both cases $\beta_3=0.$ It is remarkable that, in figure \ref{hyst1}, the quantity $<F_1^2>$ does not change at all, since the $F_1^{\mu\nu}$ does not couple to the scalar field and is not influenced by the spontaneous symmetry breaking. For the quantity $<F_2^2>$ the loop indicates a first order phase transition; the same conclusion is valid for the quantity $\cos({\rm link}),$ depicted in figure \ref{hyst2}. These statements are confirmed by long runs, which permit more accurate determination of the critical points and which clearly display two-state signals.

\begin{figure}[ht]
\begin{center}
\includegraphics[scale=0.30,angle=0]{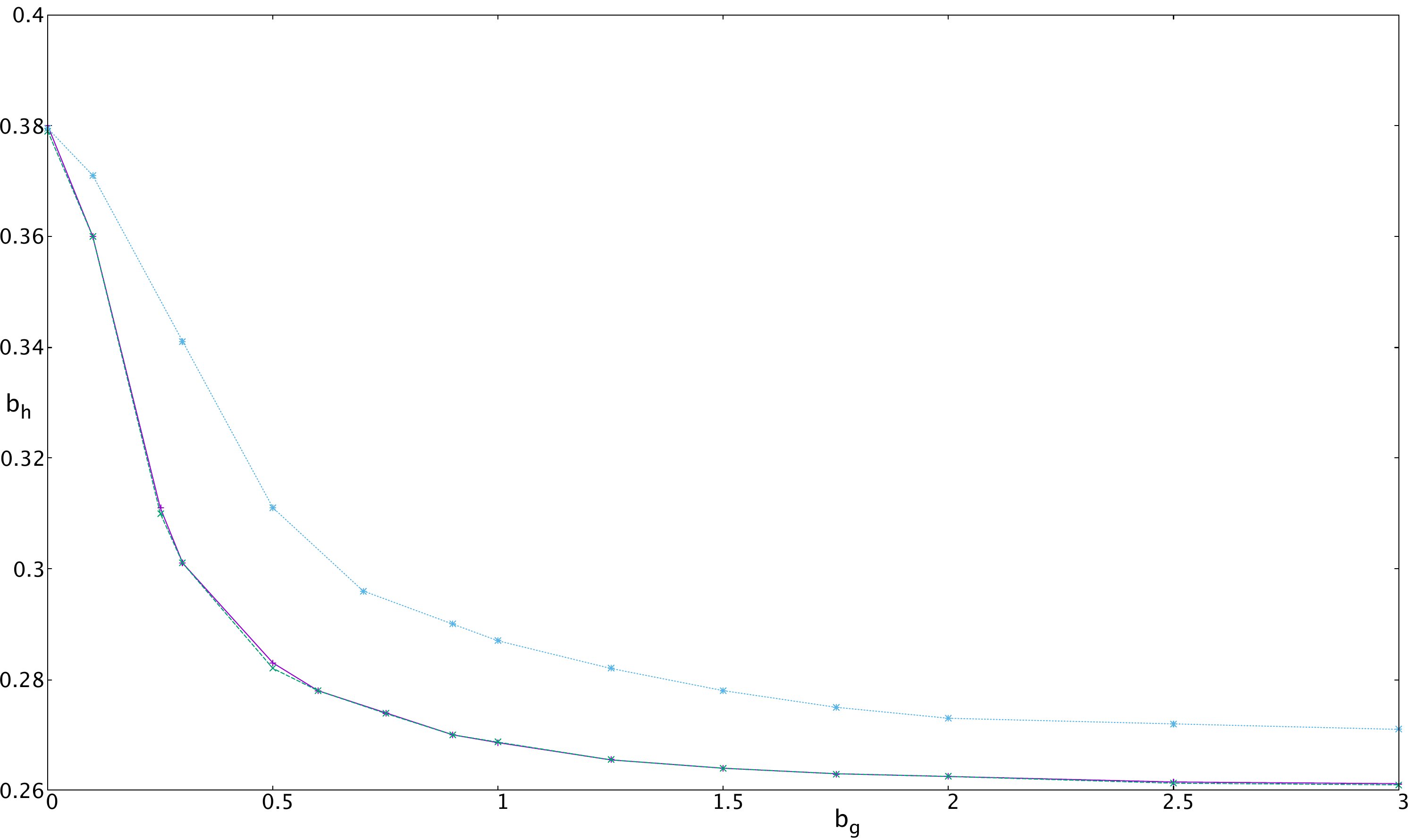}
\end{center}
\caption {Phase diagram for the three models for $\beta_3=0.00$ $(b_g\equiv \beta_g).$ Model A is represented by the uppermost curve. Notice that $\beta_g$ should be positive, otherwise the functional integral diverges.} \label{phd_ABC000}
\end{figure}

In figure \ref{phd_ABC000} we depict the phase diagrams for $\beta_3=0.00,$ using a $8^4$ lattice. It turns out that the curve for model A is very different from those for models B and C, while the latter two models almost coincide.

We depict, in figure \ref{phd_ABC060}, the phase diagram for all three models at $\beta_3=0.60,$ using once more a $8^4$ lattice. Of course, the phase diagram  makes no sense for $\beta_g < \beta_3.$ We see that the phase transition lines for model C lies at higher values of $\beta_h$ than model B, while model A lies again systematically at larger $\beta_h.$ In the latter case, model B is represented by the lower curve. It is apparent that it is possible for a given $\beta_g$ to select a value for $\beta_h,$ which places model B in the broken phase and model C in the symmetric phase. In particular, for $\beta_3=0,$ models B and C may lie in the broken phase, but, for $\beta_3>0$ there exist values for $\beta_g$ and $\beta_h,$ for which model C lies in the symmetric phase, although model B may still lie in the broken phase. This conclusion follows by inspection of figures \ref{phd_ABC000} and \ref{phd_ABC060}. In the following we concentrate on comparing B against C, since model A differs significantly from B and C, even for $\beta_3=0.$

\begin{figure}[ht]
\begin{center}
\includegraphics[scale=0.30,angle=0]{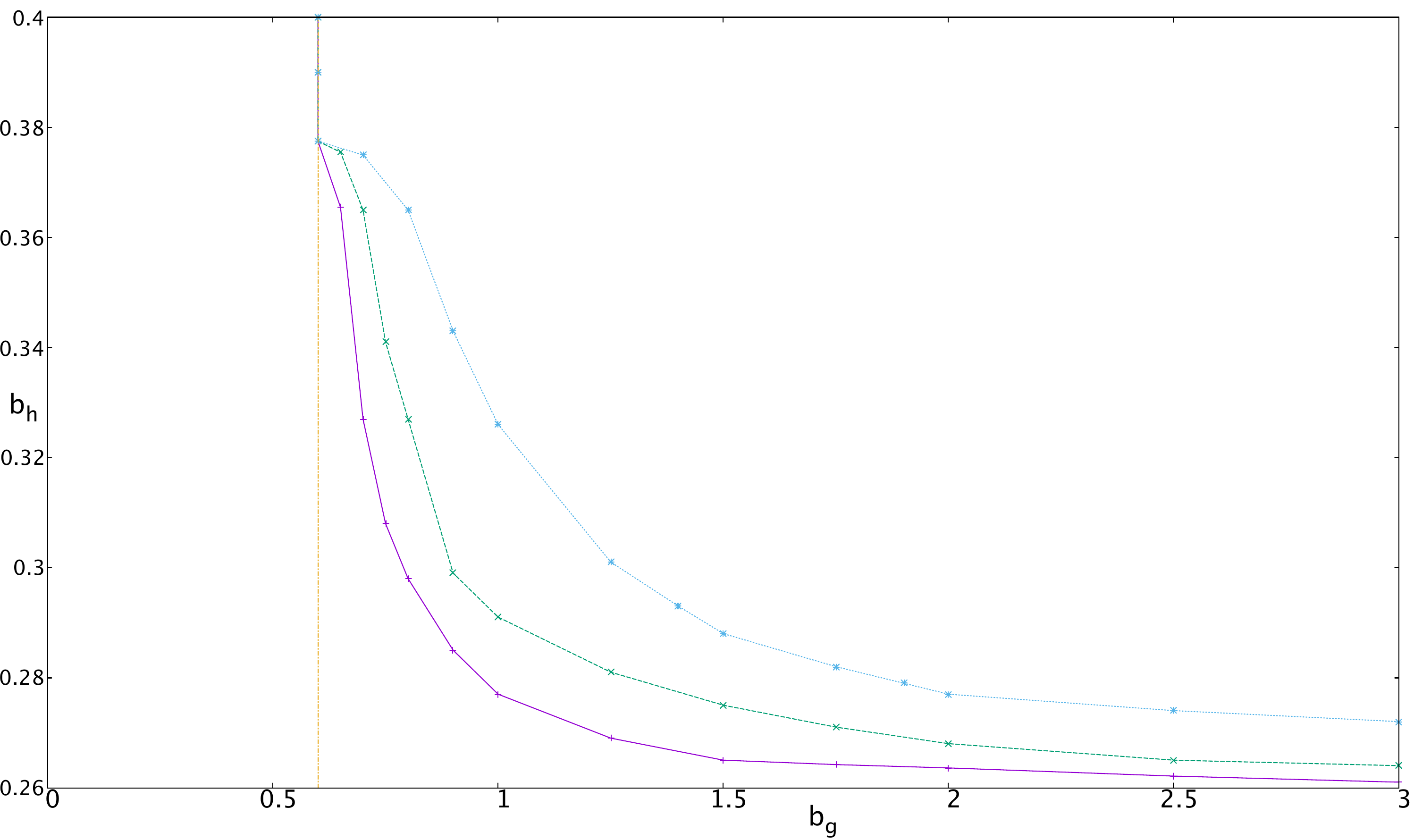}
\end{center}
\caption {Phase diagrams for models A, B and C for $\beta_3=0.60,$ represented by the vertical line. The lowest curve represents model B, the middle curve depicts model C, while the uppermost curve represents model A. The $B_u=\beta_g-\beta_3,$ which is the effective coupling of the $G_u$ field, should be positive (compare with figure \ref{phd_ABC000}); therefore the curves may not extend to the left of the vertical line. } \label{phd_ABC060}
\end{figure}

\subsection{Symmetric phase, $\beta_3$ dependence, models B and C} \label{s42}

\subsubsection{Decoupling}\label{decouple}

\begin{figure}[ht]
\begin{center}
\includegraphics[scale=0.3,angle=0]{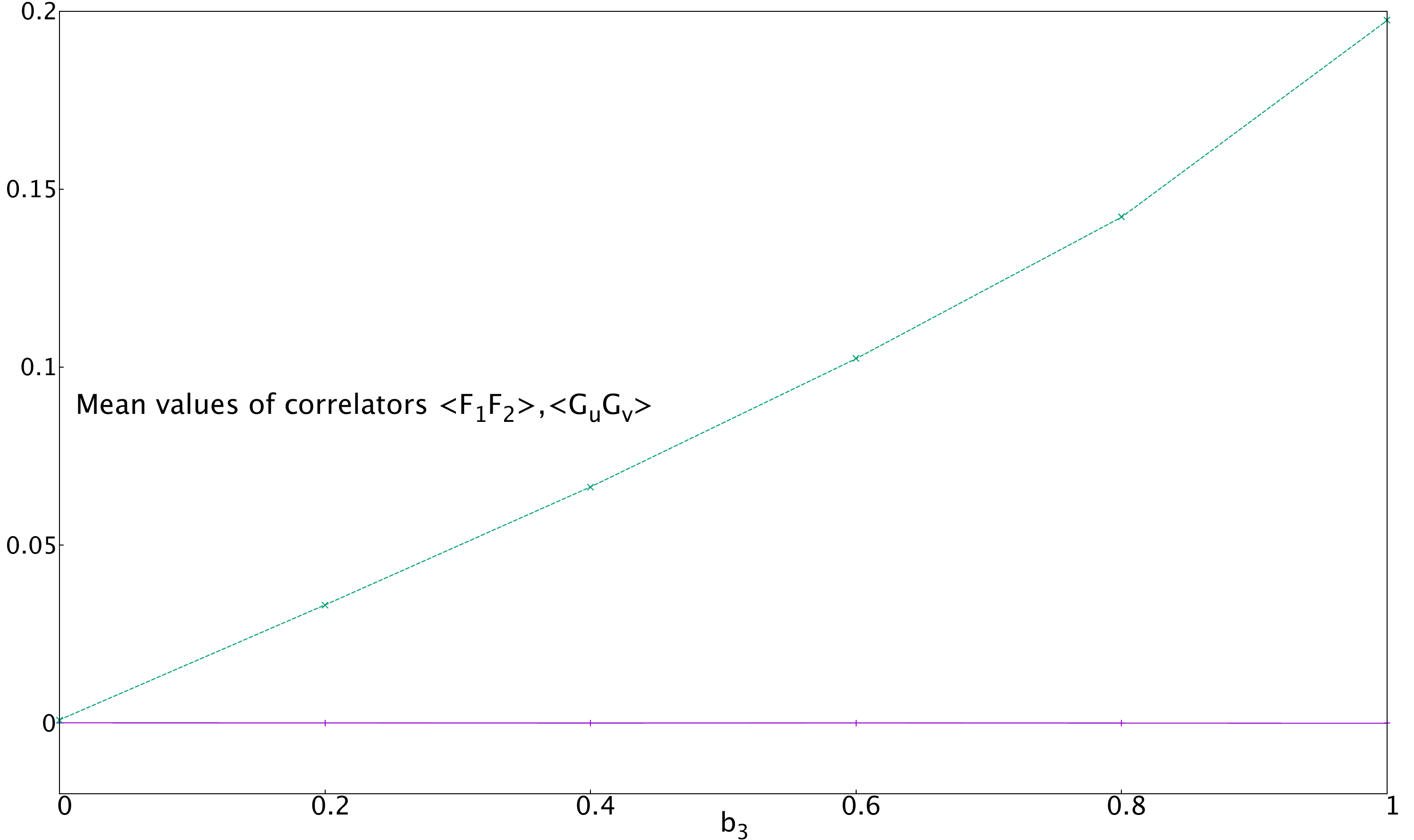}
\end{center}
\caption {Symmetric phase, correlators $< {{F}_{1}}{{F}_{2}} >$ (upper curves) and $<{{G}_{u}}{{G}_{\upsilon}}>$ (lower curves) for all models $(b_3\equiv \beta_3).$} \label{symm_corr}
\end{figure}

It is necessary to know the correlators of the various fields, so that we may calculate expectation values of fields decoupled from each other. Tree-level calculations have been described in section \ref{mixing}. The relevant correlators are given by the expressions: \be <F_1 F_2>\equiv \f{1}{6 N^4}\sum_{x}\sum_{\mu<\nu=1}^4 <F_1^{\mu\nu}(x) F_2^{\mu\nu}(x)>,\ee \be  <G_u G_v>\equiv \f{1}{6 N^4}\sum_{x}\sum_{\mu<\nu=1}^4 <G_u^{\mu\nu}(x) G_v^{\mu\nu}(x)>.\ee The results are shown in figure \ref{symm_corr}. The $<F_1 F_2>$ correlators for all models are non-zero and they diverge in the $\beta_3\to\beta_g$ limit. The important result is that the $<G_u G_v>$ correlator takes invariably a value compatible with zero, which was the point of transforming from  $F_1$ and $F_2$ to $G_u$ and $G_v$ in section \ref{mixing}. It appears that the transformation to $G_u$ and $G_v$ has been successful in decoupling the fields in the symmetric phase for all models.

Notice that \be <F_1^2> - <F_2^2> = 2 <G_u G_v>.\label{crl}\ee Since $<G_u G_v>=0,$ we predict that $<F_1^2>=<F_2^2>$ in the symmetric phase, a prediction that we have verified numerically. Notice the fact that, although we simulate the model using the $F_1$ and $F_2$ fields, the simulation automatically performs the necessary diagonalization and yields zero correlation between the $G_u$ and $G_v$ fields.

\subsubsection{Preliminary calculation of a single plaquette}

\begin{figure}[ht]
\begin{center}
\includegraphics[scale=0.30,angle=0]{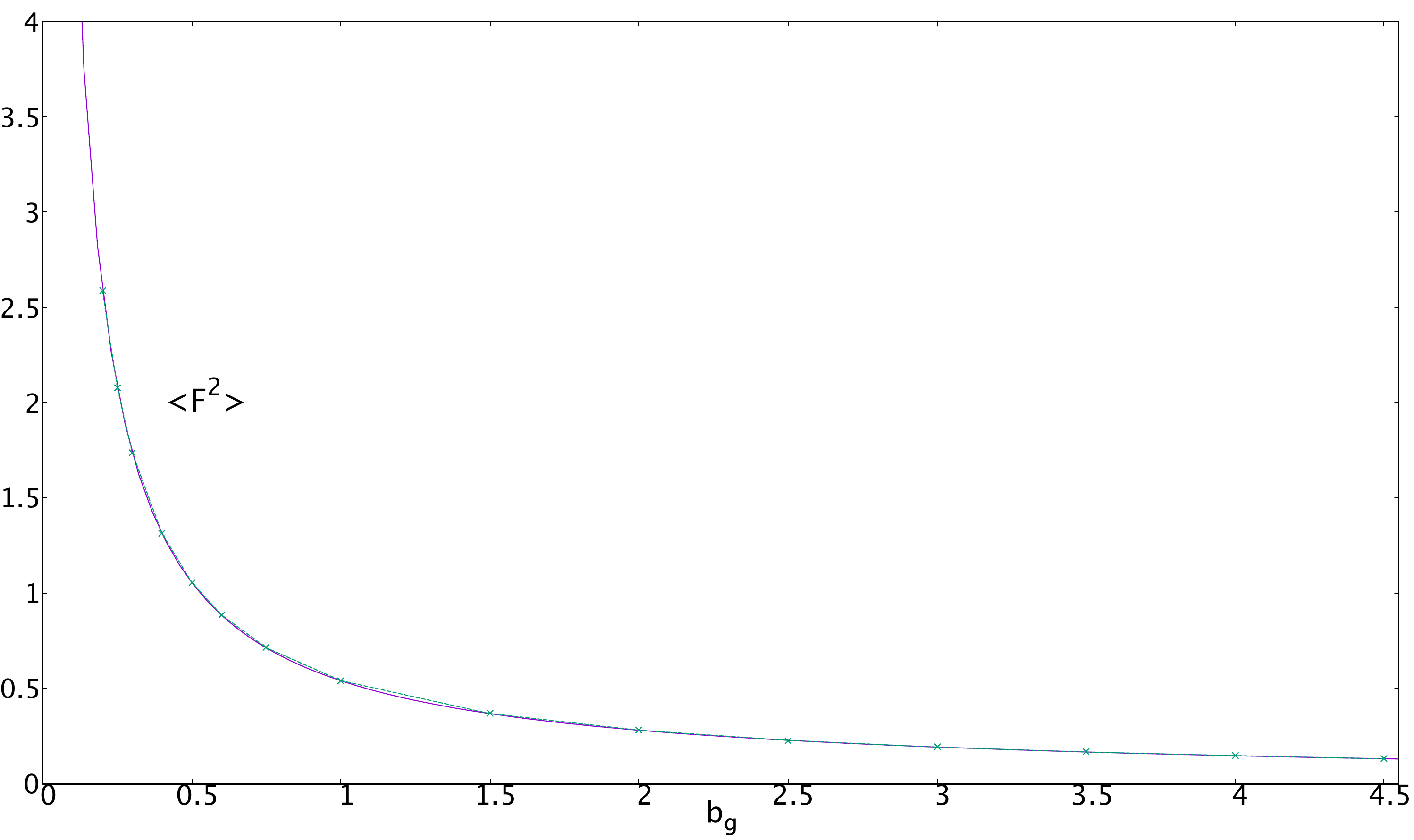}
\end{center}
\caption {Plaquette versus $\beta_g.$} \label{plbg}
\end{figure}

Another piece of information that will be useful is the expectation value of a model with just one gauge field. The action reads: \be S = \frac{1}{2} \beta_g \sum_{x,1\le\mu<\nu\le 4}F^{\mu\nu}(x)F^{\mu\nu}(x).\ee The results are depicted in figure \ref{plbg} and it turns out that the curve may be approximately represented by the expression \be <F^2> = \frac{K}{\beta_g},\  K=0.513(1),\label{singlepl}\ee for lattice volume $16^4.$ The important point is that the plaquette varies as the inverse of $\beta_g,$ a fact that will be important for us in the sequel. The parameter $K$ gets smaller values as we increase the lattice volume.

\subsubsection{$\beta_3$ dependence of the decoupled fields in the symmetric phase} \label{s423}

\begin{figure}[ht]
\begin{center}
\includegraphics[scale=0.3,angle=0]{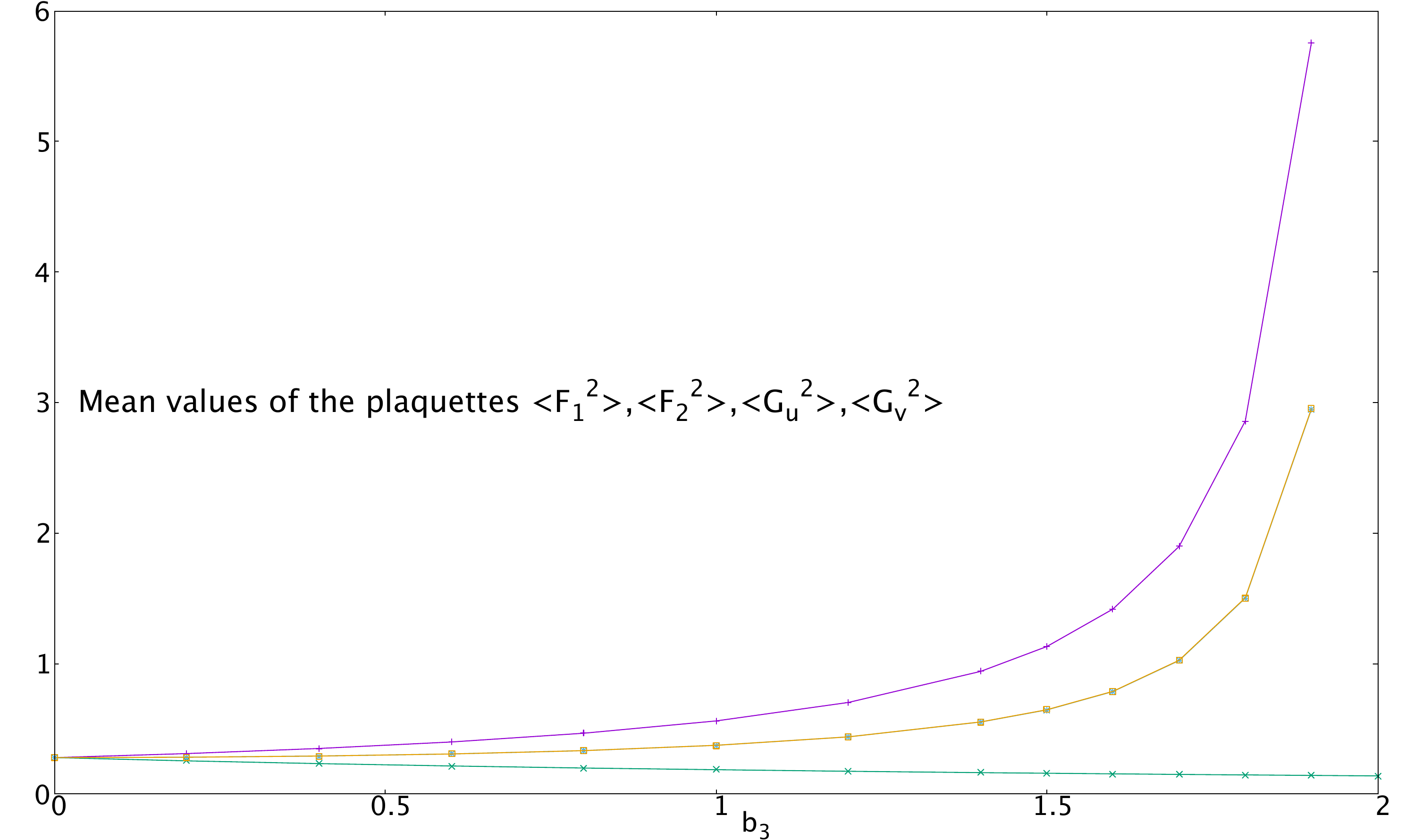}
\end{center}
\caption {Symmetric phase, $< {{F}^{2}_{1}} >$ and $<{{F}^{2}_{2}}>,$ models B and C. One may see the results for $<{{F}^{2}_{1}}>$ and $<{{F}^{2}_{2}}>,$ which are the coincident middle curve, $<G_{u}^{2}>$ (uppermost curve) and $<G_{\upsilon}^{2}>$ (lowest curve). The results are the same for all models in the symmetric phase.} \label{plaqb3asymmetric}
\end{figure}

Now we are in a position to proceed with the two-field models with a kinetic coupling term.
The picture of the symmetric phase is similar for the two models, since the symmetry breaking, which discriminates between them is absent. The parameters read: $\beta_1=\beta_2 =2.0$ and $\beta_R=0.001.$ We just consider $\beta_3>0.$ The quantities $<F_1^2>$ and $<F_2^2>$ are shown in figure \ref{plaqb3asymmetric}, for the symmetric phase for models B and C (where $\beta_h=0.23$). The two quantities are represented by the middle curve and they are equal (this agrees with the prediction based on the vanishing of the correlators, already mentioned previously, equation (\ref{crl})). The quantities  $<F_1^2>$ and $<F_2^2>$ start from moderate values and increase with $\beta_3.$ However, we recall that the $F_1$ and $F_2$ fields are highly correlated, due to the kinetic mixing term, so one should rather consider the fields $G_u$ and $G_v,$ which are uncorrelated in the symmetric phase, as seen above. Since the two fields are uncorrelated, we expect to get results which will behave as suggested by (\ref{singlepl}), their only difference being that $<G_u^2>$ is inversely proportional to $B_u=\beta_g-\beta_3,$ while $<G_v^2>$ is inversely proportional to $B_v=\beta_g+\beta_3.$ In figure \ref{plaqb3asymmetric} one may see the results for $<G_u^2>$ (it is the uppermost curve) and we may inspect the divergence of this quantity as $\beta_g \to \beta_3 \ (B_u\to 0).$  On the other hand, $<G_v^2>$ (the lowest curve) decreases with increasing $\beta_3,$ since $\beta_g+\beta_3$ never vanishes.

We now proceed with fitting the expectation values of $<G_u^2>$ and $<G_v^2>$ with $\f{1}{\beta_g-\beta_3}$ and $\f{1}{\beta_g+\beta_3}$ respectively. The fits yield: \be <G_u^2> = \f{0.569(1)}{\beta_g-\beta_3},\ <G_v^2> = \f{0.564(1)}{\beta_g+\beta_3}. \ee We observe that the numerators are somewhat different from $0.513(1)$ found from the single gauge field before, equation (\ref{singlepl}). It means that the system, although it is similar to a set of independent gauge fields $A_u,\ A_v,$ it is not entirely the same, since interaction with the scalar field is also involved.

We have also calculated the Wilson loops for $G_u$ and $G_v$ and we have fit them using the expressions: \be W_u(R,T)\propto \exp\left[-\f{\alpha(G_u)}{R} T\right],\ \ W_v(R,T)\propto \exp\left[-\f{\alpha(G_v)}{R} T\right].\ee The results are shown in table 1.

We may calculate the tree-level fine structure constant using the relations $\alpha = \f{g^2}{4 \pi}$ and $\beta_g=\f{1}{g^2}.$ If we eliminate $g^2$ among the two expressions, we get $\alpha = \f{1}{4 \pi \beta_g}.$ One may compare the value of $\alpha$ from the Wilson loops against the tree-level prediction, after substituting $\beta_g$ with the quantities $B_u = \beta_g - \beta_3,\ B_v = \beta_g + \beta_3,$ since these are the expected couplings for $G_u,\ G_v.$

We observe that the fine structure constants calculated from the Wilson loops are in general larger than the tree level values. However, the fine structure constants follow the variation of the tree level parameters. The coupling $\alpha(G_u)$ increases, while $\alpha(G_v)$ decreases with $\beta_3.$

We conclude that $\alpha(G_u)$ differs from $\alpha(G_v),$ just because the two  fields $G_u$ and $G_v$ depend on $B_u$ and $B_v$ respectively. There is no important difference between models B and C in the symmetric phase. One may fit the data and discover that: \be \alpha(G_u) \simeq \f{1.18}{4 \pi (\beta_g-\beta_3)},\ \ \alpha(G_v) \simeq \f{1.18}{4 \pi (\beta_g+\beta_3)}.\ee

$$
    \begin{tabular} { | c | c | c | c | c | c | c |}
        \hline
        $\beta_3$ &  $\f{1}{4 \pi (\beta_g-\beta_3)}$  & $\alpha(G_u)$  &  $\f{1}{4 \pi (\beta_g+\beta_3)}$ &    $\alpha(G_v)$  \\
    \hline
        $0.00$    &  $0.0398$  & $0.0469(4)$  &  $0.0398$  & $0.0470(3)$  \\
       \hline
        $0.05$    &  $0.0408$  & $0.0483(4)$  &  $0.0388$  & $0.0461(3)$  \\
        \hline
        $0.10$    &  $0.0419$  & $0.0494(5)$  &  $0.0379$  & $0.0445(4)$   \\
        \hline
        $0.20$    &  $0.0442$  & $0.0522(5)$  &  $0.0362$  & $0.0426(4)$   \\
        \hline
        $0.30$    &  $0.0468$  & $0.0550(5)$  &  $0.0346$  & $0.0411(4)$   \\
        \hline
        $0.40$    &  $0.0497$  & $0.0587(5)$  &  $0.0332$  & $0.0392(4)$   \\
        \hline
        $0.50$    &  $0.0531$  & $0.0627(5)$  &  $0.0318$  & $0.0377(2)$   \\
        \hline
        $0.60$    &  $0.0569$  & $0.0682(7)$  &  $0.0306$  & $0.0352(4)$   \\
        \hline
        $0.70$    &  $0.0612$  & $0.0730(6)$  &  $0.0295$  & $0.0333(4)$   \\
        \hline
        $0.80$    &  $0.0663$  & $0.0791(7)$  &  $0.0284$  & $0.0314(4)$   \\
        \hline
        $0.90$    &  $0.0723$  & $0.0857(7)$  &  $0.0274$  & $0.0294(4)$   \\
        \hline
        $1.00$    &  $0.0796$  & $0.0908(7)$  &  $0.0265$  & $0.0276(5)$   \\
        \hline
    \end{tabular}
$$
Table 1: Predictions and results for all models in the symmetric phase.

\subsection{Broken Phase: Model B $(q_1=0,\ q_2=1)$} \label{s43}

\begin{figure}[ht]
\begin{center}
\includegraphics[scale=0.3,angle=0]{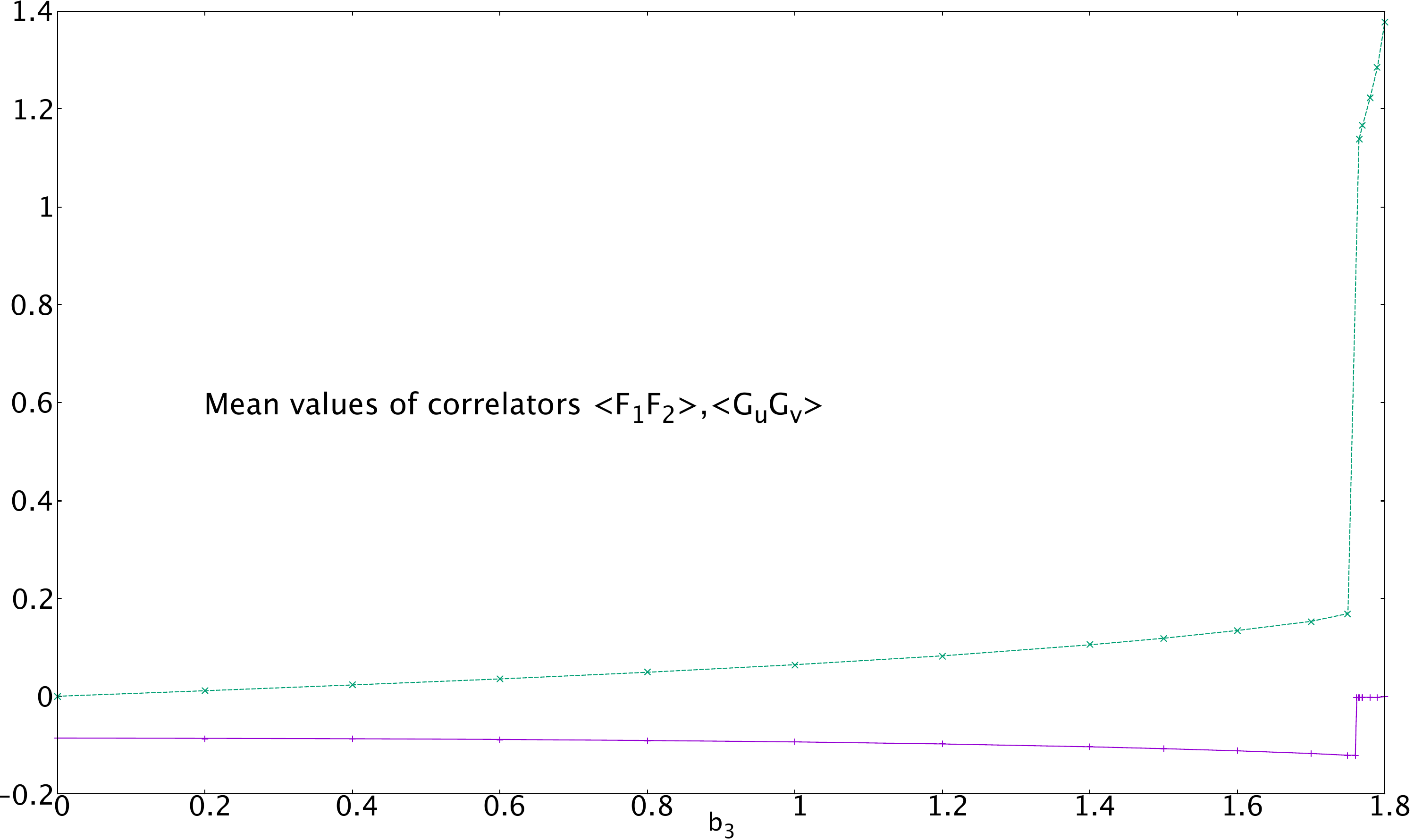}
\end{center}
\caption {Broken phase, model B: $<F_1F_2>$ correlators (upper curves)  and $<G_uG_\upsilon>$ correlators (lower curves). The $<F_1F_2>$ correlators are greater than zero for the most part, while the $<G_uG_\upsilon>$ correlators are mostly smaller than zero. When $\beta_3$ becomes greater than $1.75,$ the system passes over to the symmetric phase and the $G_u$ and $G_\upsilon$ fields are the appropriate fields for its description; this means that they are decorrelated and this is why their correlator becomes zero for these $\beta_3$ values.} \label{correlB}
\end{figure}

We set $\beta_h=0.29,$ which places the system in the broken  phase. We start with the dependence of various quantities on $\beta_3$. Figure \ref{correlB} shows that neither $F_1,\ F_2,$  nor $G_u,\ G_v$ are appropriate for the study of the model, since both correlators $<F_1 F_2>$ and  $<G_u G_v>$ are different from zero (with the exception of $<F_1 F_2>$ at $\beta_3=0).$ There is a striking feature of these diagrams that needs explanation, namely the fact that the $<F_1 F_2>$  correlators appear to take very large values for $\beta_3\ge 1.75.$ We shall deal with this later on. We have restricted plotting to $0\le \beta_3\le 1.80,$ since the values of $<F_1 F_2>$ for larger $\beta_3$ are too large to depict in the graph. Another remarkable feature in the graph is the fact that the correlator $<G_u G_v>$ vanishes for $\beta_3\in (1.75,\ 1.80).$ We will also come to this later on.

\begin{figure}[ht]
\begin{center}
\includegraphics[scale=0.3,angle=0]{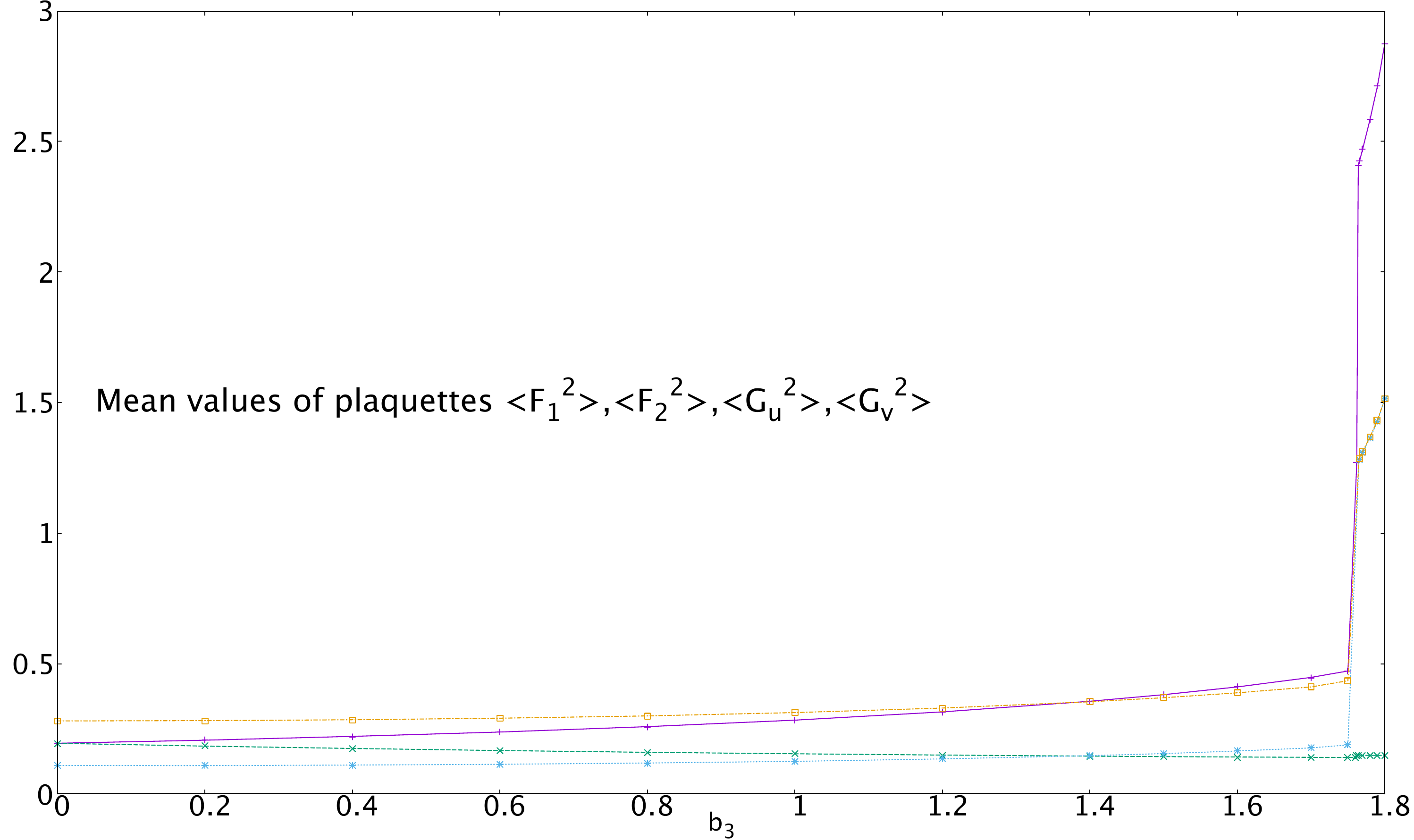}
\end{center}
\caption {Broken phase, Model B : $<G^{2}_u >$ and $<G^{2}_\upsilon>$ start together and diverge gradually with increasing $\beta_{3}$. $<G^{2}_u >$ is the one with bigger values. The remaining curves represent $<F^{2}_1>$ (the curve with the larger values, taking the value 0.28 for $\beta_3=0.$) and $<F^{2}_2>$ (the curve with the smaller values, starting at 0.11).} \label{plaquetteb3B}
\end{figure}

We calculated the plaquettes, which will prove useful in the sequel. The results for $<F_1^2>$ and $<F_2^2>$ (figure \ref{plaquetteb3B}), are not equal any more: the plaquette $<F_1^2>$ of the visible field starts from a fairly large value, since it is in the unbroken phase, while the dark field $<F_2^2>$ starts at small value, since it lies in the broken phase; both increase with $\beta_3.$ These results are consistent with the ones in figure \ref{hyst1}, where the $<F_1^2>$ and the $<F_2^2>$ loops are shown. On the other hand, $<G_u^2>$ increases, while $<G_v^2>$ slightly decreases with $\beta_3.$ It appears that the pair $F_1^{\mu\nu},\ F_2^{\mu\nu}$ behaves as expected from the tree level calculations: $<F_1^2>$ takes a relatively large value, since it corresponds to an unbroken component, while $<F_2^2>$ takes small values, compatible with a broken phase observable. On the other hand the plaquette values, with the interesting exception of $<G^{2}_\upsilon>,$ are also taking very large values for $\beta_3\ge 1.75,$ much like the correlators in figure \ref{correlB}.

\begin{figure}[ht]
\begin{center}
\includegraphics[scale=0.30,angle=0]{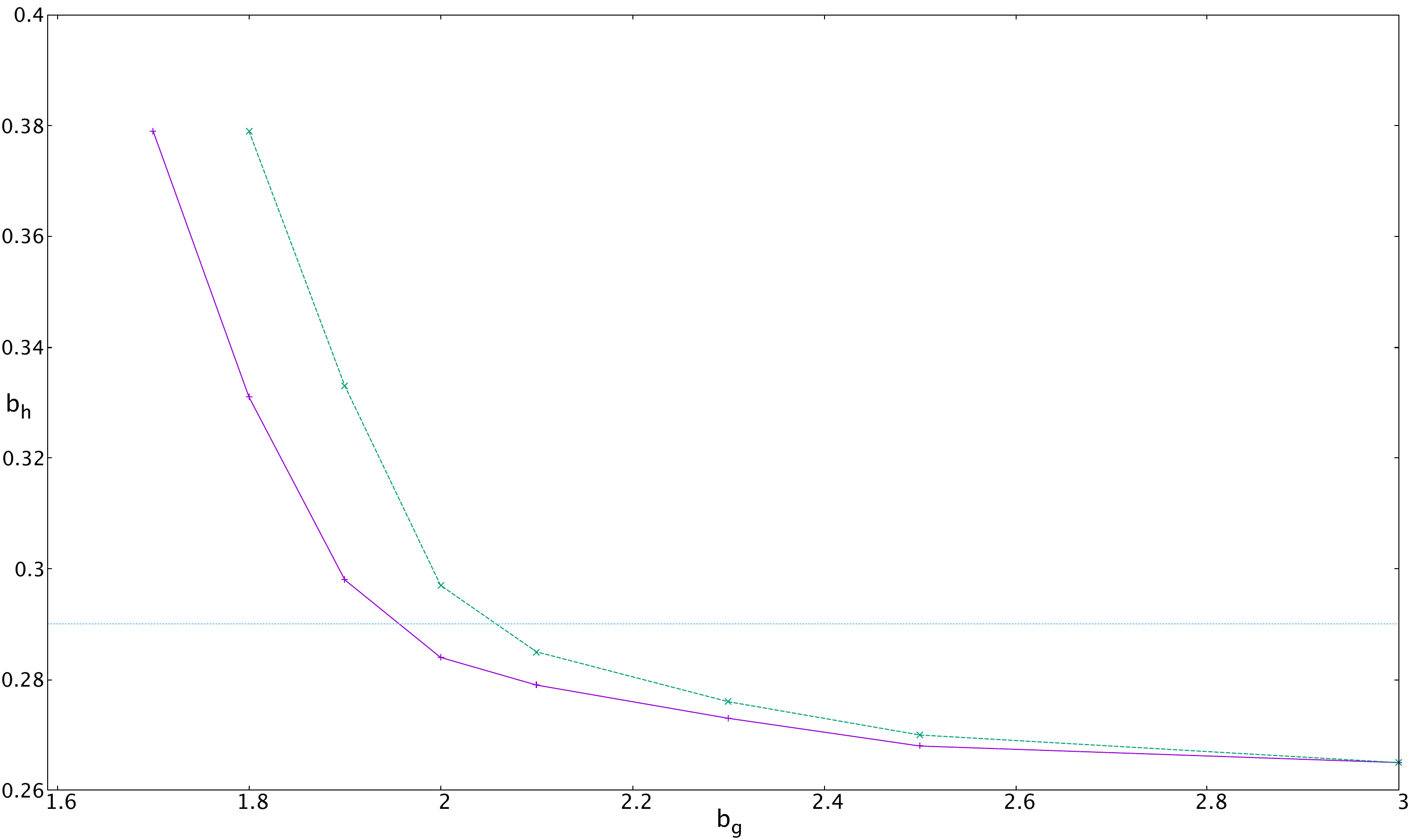}
\end{center}
\caption {Broken phase, model B: Phase diagram for $\beta_{3}=1.70$ (lower curve) $\beta_{3}=1.80$ (upper curve). The horizontal line represents the $\beta_h=0.29$ value used in the simulations.} \label{fig:phd170_180}
\end{figure}

To understand the abrupt change in the behaviours of the $<G_uG_\upsilon>$ correlators in figure \ref{correlB}, as well as of the $<G^{2}_u >,\ <G^{2}_v>$ and $<F_1^2>,\ <F_2^2>$ plaquettes in figure \ref{plaquetteb3B}, we depict in figure \ref{fig:phd170_180} the phase diagram for the two values $\beta_{3}=1.70$ and $\beta_{3}=1.80.$ If we concentrate in $\beta_g=2.0,\ \beta_h=0.29,$ i.e. the values used throughout this work, it is seen that the system is in the broken phase for $\beta_3=1.70$ (lower curve), and in the symmetric phase for $\beta_3=1.80$ (upper curve). Thus the large values seen before in the plaquettes are connected with the passage of the system to the symmetric phase. The same can be said about the $<G_uG_\upsilon>$ correlator, which vanishes abruptly for $\beta_3\ge 1.75:$ this may also be attributed to the passage to the symmetric phase. In addition, the  plaquette $<G_v^2>$ takes the symmetric phase value in this parameter region.

As explained above, the problem with the results so far is that the field correlators $<F_1 F_2>,$ as well as $<G_u G_v>,$ don't vanish, that is the corresponding fields exhibit a non-trivial mixing. It is desirable to find linear combinations of the fields, which yield uncorrelated fields. We rotate the $F_1^{\mu\nu},\ F_2^{\mu\nu}$ fields to $H_1^{\mu\nu},\ H_2^{\mu\nu}$ through an angle $\varphi:$ \be H_1^{\mu\nu} = \sin\varphi F_1^{\mu\nu} +\cos \varphi F_2^{\mu\nu},\ \ H_2^{\mu\nu} = \cos \varphi F_1^{\mu\nu} -\sin\varphi F_2^{\mu\nu},\ee and vary this angle until the correlators $<H_1^{\mu\nu}H_2^{\mu\nu}>$ of the resulting fields $H_1^{\mu\nu},\ H_2^{\mu\nu},$ vanish. The fields $H_1^{\mu\nu}$ and $H_2^{\mu\nu}$ are the lattice analogs of the fields ${\cal X}_1^{\mu\nu}$ and ${\cal X}_2^{\mu\nu}$ mentioned in the continuum treatment of the model. Some interesting special values for $\phi$ are:

(a) $\varphi=\f{\pi}{2}\simeq 1.5708,$ for which $H_1^{\mu\nu} = F_1^{\mu\nu},\ H_2^{\mu\nu}=-F_2^{\mu\nu}.$ At $\beta_3=0$ the $H_1^{\mu\nu}$ and $H_2^{\mu\nu}$ fields are exactly equal to $F_1^{\mu\nu}$ and $-F_2^{\mu\nu}$ respectively, while for $\beta_3>0$ these expressions also involve admixtures of  $F_2^{\mu\nu}$ and  $F_1^{\mu\nu}$ respectively (see table 2).

(b) $\varphi=\f{\pi}{4}\simeq 0.7854,$ for which $H_1^{\mu\nu} = \f{F_1^{\mu\nu}+F_2^{\mu\nu}}{\sqrt{2}}=G_u^{\mu\nu},\ H_2^{\mu\nu}=\f{F_1^{\mu\nu}-F_2^{\mu\nu}}{\sqrt{2}}=-G_v^{\mu\nu},\ \ $ that is the linear combinations appropriate for models A and C.

For small $\beta_3$ we expect that $H_1^{\mu\nu}$ will be approximately equal to $F_1^{\mu\nu},$ as we saw in our analysis in the paragraph about matter fields. By the same analysis, $H_2^{\mu\nu}$ will be proportional to $F_2^{\mu\nu}$ (equation (\ref{Bx1x2}) of subsection \ref{s22}). The relevant angles for selected values of $\beta_3$ are shown in table 2. One may see that the appropriate fields, $H_1$ and $H_2,$ are not exactly equal to $F_1$ and $-F_2,$ as the tree level considerations suggest, however they are mainly $F_1$ (with small $F_2$ admixture) and $-F_2$ (with small $F_1$ admixture) for a quite large range of $\beta_3,$ say up to $\beta_3\simeq 0.30.$ For bigger $\beta_3$ the contributions of $F_1$ in $H_2$ and the one of $F_2$ in $H_1$ become significant.

\begin{figure}[ht]
\begin{center}
\includegraphics[scale=0.3,angle=0]{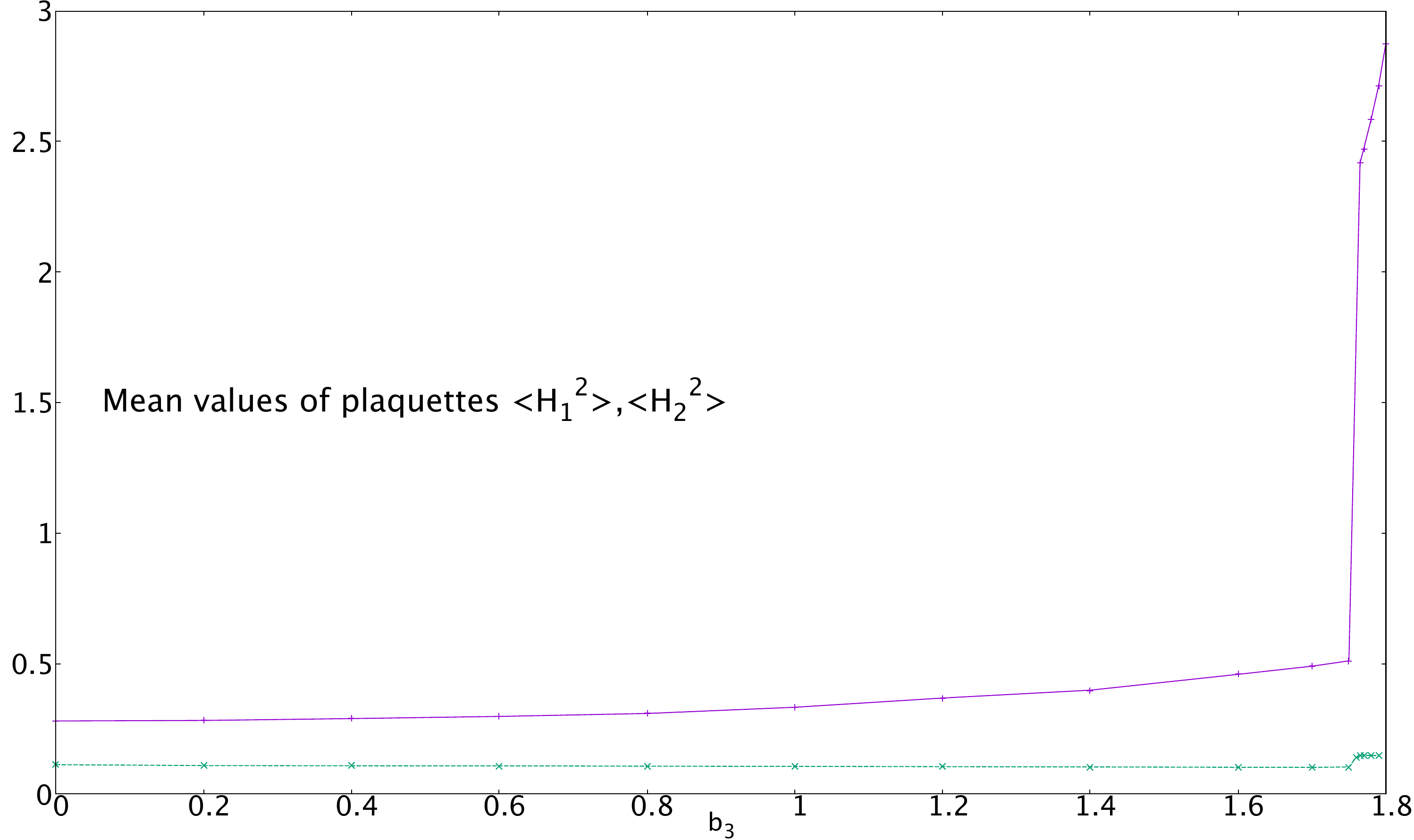}
\end{center}
\caption {Broken phase, model B: $<H^{2}_1>$ (upper curve) increases with increasing $\beta_{3}$ while  $<H^{2}_2>$ (lower curve) remains approximately constant (actually it decreases slightly) until $\beta_3$ takes the value 1.75.} \label{fig:huhv}
\end{figure}

In figure \ref{fig:huhv} we show the results of the simulation for $<H_1^2>$ and $<H_2^2>$ versus $\beta_3.$ It appears that $<H_2^2>$ is approximately constant (slightly decreasing), while $<H_1^2>$ increases slightly as $\beta_3$ increases. This is to be expected, since  $<H_1^2>$ equals mostly $<F_1^2>,$ which does not couple to the scalar field, while $<H_2^2>$ equals mostly $<F_2^2>,$ so that it is influenced very much by the symmetry breaking. As $\beta_3$ increases and $\varphi$ tends to $\f{\pi}{4},$ $H_1^{\mu\nu}$ approaches $G_u^{\mu\nu},$ while $H_2^{\mu\nu}$ approaches $G_v^{\mu\nu},$ the fields that mostly  characterize the symmetric phase. This transition is shown in figure \ref{fig:huhv} for $\beta_3$ around 1.75.

$$
    \begin{tabular} { | c | c | c |c | }
        \hline
        $\beta_3$ &  $\varphi$ &  $H_1$ &  $H_2$ \\
    \hline
        $0.00$    &      $1.5708$ &      $F_1$&      $-F_2$ \\
    \hline
        $0.10$    &      $1.5381$ &    $0.999 F_1 + 0.033 F_2$ &  $0.033 F_1 - 0.999 F_2$       \\
    \hline
        $0.20$    &      $1.5053$ &    $0.998 F_1 + 0.065 F_2$ &  $0.065 F_1 - 0.998 F_2$       \\
    \hline
        $0.30$    &      $1.4730$ &    $0.995 F_1 + 0.098 F_2$ &  $0.098 F_1 - 0.995 F_2$      \\
     \hline
        $0.40$    &      $1.4408$ &    $0.992 F_1 + 0.130 F_2$ &  $0.130 F_1 - 0.992 F_2$      \\
        \hline
        $0.50$    &      $1.4097$ &    $0.987 F_1 + 0.160 F_2$ & $0.160 F_1 - 0.987 F_2$      \\
   \hline
        $0.60$    &      $1.3793$ &    $0.982 F_1 + 0.190 F_2$ &  $0.190 F_1 - 0.982 F_2$       \\
   \hline
        $0.70$    &      $1.3430$ &    $0.974 F_1 + 0.226 F_2$ &  $0.226 F_1 - 0.974 F_2$       \\
     \hline
        $0.80$    &      $1.3200$ &    $0.969 F_1 + 0.248 F_2$ &  $0.248 F_1 - 0.969 F_2$       \\
        \hline
        $0.90$    &      $1.2925$ &    $0.962 F_1 + 0.275 F_2$ &  $0.275 F_1 - 0.962 F_2$       \\
   \hline
        $1.00$    &      $1.2625$ &    $0.953 F_1 + 0.303 F_2$ &  $0.303 F_1 - 0.953 F_2$       \\
   \hline
\end{tabular}
$$
Table 2: Rotation angles for the construction of uncorrelated fields $H_1,\ H_2,$ in the broken phase of model B.

$$
    \begin{tabular} { | c | c | c | c | c | c | c | }
    \hline
        $\beta_3$ &  $\varphi$ & $m(F_2)$ & $\alpha(F_1)$ &  $\alpha(H_1)$ & $m(H_2)$ &  $m_{scalar}$ \\
    \hline
        $0.00$    &   $1.5708$ &  $2.15(4)$ &   $0.0470(2)$ &    $0.0470(2)$ & $2.15(4)$ & $1.91(7)$ \\
    \hline
        $0.10$    &   $1.5381$ &  $2.31(3)$ &   $0.0469(3)$ &    $0.0470(2)$ & $1.905(6)$ & $1.92(6)$\\
    \hline
        $0.20$    &   $1.5053$ &  $2.32(4)$ &   $0.0468(2)$ &    $0.0471(1)$ & $1.65(9)$ & $1.93(7)$ \\
    \hline
        $0.30$    &   $1.4730$ &  $2.31(3)$ &   $0.0471(2)$ &     $0.0469(3)$ & $1.52(9)$ & $1.93(8)$ \\
    \hline
        $0.40$    &   $1.4408$ &  $2.24(4)$ &   $0.0472(2)$ &    $0.0470(1)$ & $1.48(11)$ & $1.94(8)$ \\
    \hline
        $0.50$    &   $1.4097$ &  $2.27(2)$ &   $0.0469(2)$ &    $0.0468(3)$ & $1.30(9)$ &  $1.95(6)$ \\
    \hline
        $0.60$    &   $1.3793$ &  $2.36(4)$ &   $0.0474(3)$ &    $0.0470(2)$ & $1.23(10)$ & $1.95(7)$ \\
    \hline
        $0.70$    &   $1.3430$ &  $2.28(2)$ &   $0.0469(2)$ &    $0.0460(3)$ & $1.10(10)$ & $1.97(7)$ \\
    \hline
        $0.80$    &   $1.3200$ &  $2.24(4)$ &   $0.0475(2)$ &     $0.0460(2)$ & $0.94(8)$ & $1.99(6)$ \\
    \hline
        $0.90$    &   $1.2925$ &  $2.26(3)$ &   $0.0474(2)$ &    $0.0450(2)$ & $0.87(2)$ &  $2.00(7)$ \\
    \hline
        $1.00$    &   $1.2625$ &  $2.21(7)$ &   $0.0473(2)$ &     $0.0453(2)$ & $0.69(9)$ &  $2.02(6)$ \\
    \hline
\end{tabular}
$$

Table 3: Fit parameters for the fields $F_1^{\mu\nu}$ and $F_2^{\mu\nu},$ as well as the uncorrelated fields $H_1^{\mu\nu}$ and $ H_2^{\mu\nu},$ in the broken phase of model B.

We have also calculated the Wilson loops for $H_1$ and $H_2$ and fit them using the expressions:
\be W_{F_2}(R,T)\propto \exp\left[-\f{\exp\left[-m(F_2) R\right]}{R} T \right],\ \ W_{H_2}(R,T)\propto \exp\left[-\f{\exp\left[-m(H_2) R\right]}{R} T \right] \ee \be W_{F_1}(R,T)\propto \exp\left[-\f{\alpha(F_1)}{R} T\right],\ \ W_{H_1}(R,T)\propto \exp\left[-\f{\alpha(H_1)}{R} T\right].\ee The results are shown in table 3. In fact, we have proceeded up to $\beta_3=1.80$ (not shown in the table) and found that $m(H_2)$ is compatible with zero, which agrees with the previous conclusion that the system passes to the symmetric phase for these values of $\beta_3.$ The value of $\alpha(H_1)$ does not change much. It is interesting that $\alpha(H_1)$ is approximately constant and equal to $\alpha(F_1),$ which is expected, since this field is presumably not coupled to the scalar field.

We see that, for $\beta_3=0,$ $H_2^{\mu\nu}$ coincides with the $-F_2^{\mu\nu}$ field, which interacts with the scalar field and is best simulated with a Yukawa form for the potential in the broken phase; on the contrary, for $\beta_3=0$ again, $H_1^{\mu\nu}$ coincides with the $F_1^{\mu\nu}$ field, which does not interact  with the scalar field, so one should expect the behaviour appropriate to pure gauge fields. For non-zero $\beta_3$ there is a difference, in the sense that it is necessary to include a slight admixture of $F_1^{\mu\nu}$ into $H_2^{\mu\nu},$ apart from the $F_2^{\mu\nu}.$ This is not predicted by the tree level considerations. Let us recall equations (\ref{Bx1x2}):
\be {\cal X}_1^\mu = -{\cal A}_1^\mu+\kappa{\cal A}_2^\mu,\ \ \ {\cal X}_2^\mu = \sqrt{1-\kappa^2} {\cal A}_2^\mu.\ee
The tree-level field $\p_\mu {\cal X}_{1, \nu} - \p_\nu {\cal X}_{1, \mu}$ corresponds to $H_{1,\mu\nu},$ while $\p_\mu {\cal X}_{2, \nu} - \p_\nu {\cal X}_{2, \mu}$ corresponds to $H_{2,\mu\nu}.$ We find that, for small $\kappa,$ ${\cal X}_1^\mu$ derives from ${\cal A}_1^\mu$ only, and this is consistent with the results of table 2, where the corresponding field $H_1^{\mu\nu}$ is basically $F_1^{\mu\nu}$ (with an admixture of $F_2^{\mu\nu},$ however).
On the lattice we need an admixture of $F_1^{\mu\nu}$ and $F_2^{\mu\nu}$ to get the uncorrelated fields $H_1^{\mu\nu}$ and $H_2^{\mu\nu}.$ Thus, for small $\beta_3,$ it turns out that $H_1$ is basically $F_1$ and yields  a Coulomb potential, while $H_2$ is basically $F_2$ and  yields a Yukawa potential. It is reasonable to expect Coulomb expressions for the $H_1$ fields and Yukawa expressions for the $H_2$ fields; the numerical results confirm these expectations. Having fixed the linear transformations for each value of $\beta_3,$ we calculate the potentials for $H_1$ and $H_2$ and fit $H_2$ with a Yukawa form with mass $m(H_2)$ and $H_1$ with a Coulomb form, where the fine structure is $\alpha(H_1).$

We have also reported, in table 3, the masses of the scalar field, to compare with the mass of the $H_2$ gauge field. It turns out that the mass of the scalar field is approximately independent from $\beta_3,$ while $m(H_2)$ is comparable to $m_{scalar}$ for small $\beta_3,$ but it drops substantially when $\beta_3$ increases. The scalar field can possibly decay into the gauge $H_2$ and other fields, while the products of the $H_2$ decay may not include the scalar, especially for $\beta_3 >0.10.$

\subsection{Broken Phase: Model C $\left(q_1=\f{1}{\sqrt{2}},\ q_2= \f{1}{\sqrt{2}}\right)$} \label{s44}

\begin{figure}[ht]
\begin{center}
\includegraphics[scale=0.3,angle=0]{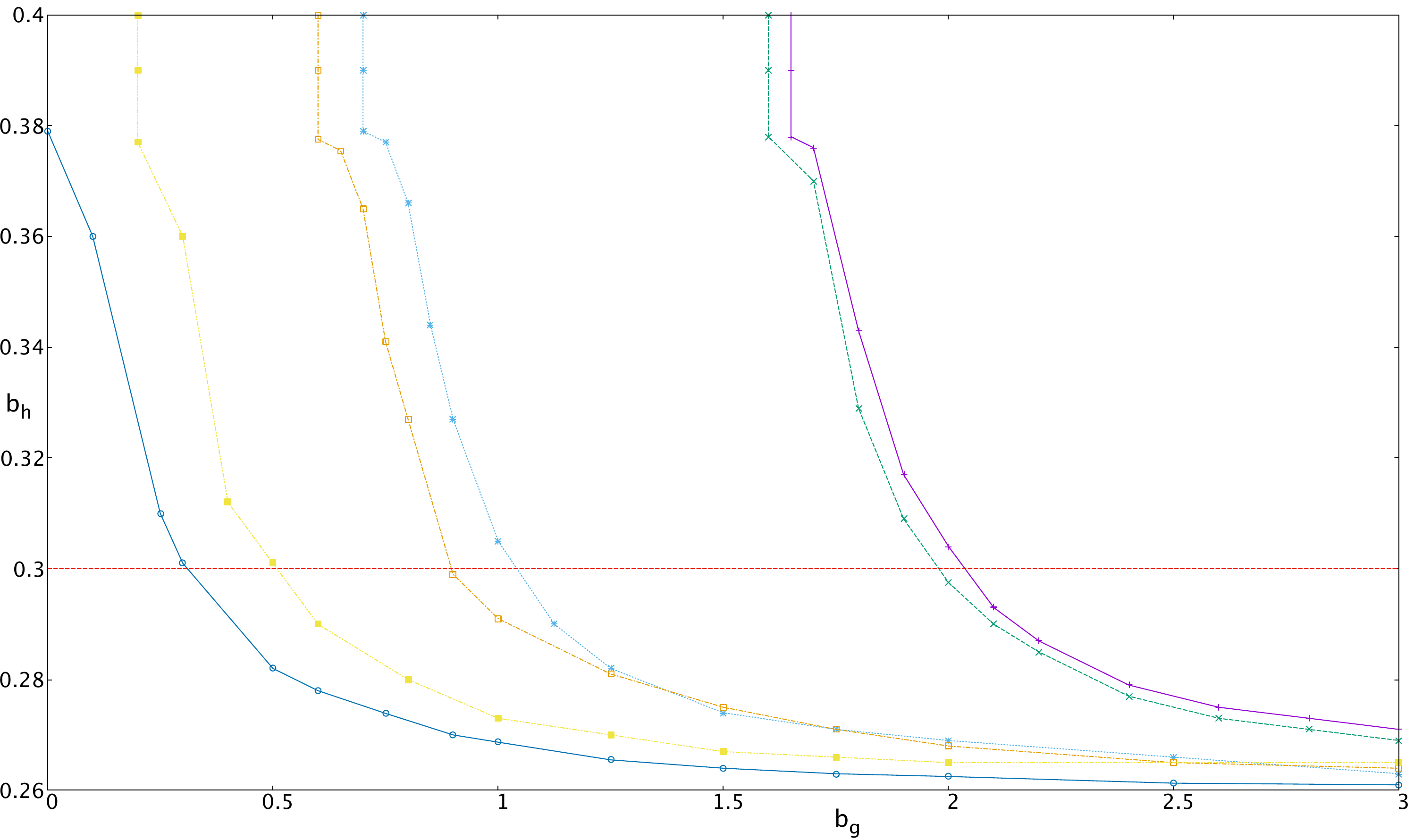}
\end{center}
\caption {Phase diagrams for model C and the values: $\beta_{3}=0.0,\ \beta_{3}=0.20$ (the leftmost curves), $ \beta_{3}=0.60,\ \beta_{3}=0.70$ (the two curves in the middle) and $\beta_{3}=1.60,\ \beta_{3}=1.65$ (the two rightmost curves).} \label{C_all}
\end{figure}

We start our study of model C with its phase diagram in figure \ref{C_all}. In this figure we show the phase diagrams for model C at six values of $\beta_3,$ namely $0.00,\, 0.20,\ 0.60,\ 0.70,\ 1.60$ and $1.65.$ If we focus on $\beta_g=2.0,$ the value used throughout our paper, we observe that, for $\beta_{3}=1.60,\ \beta_h=0.30,$ the system lies in the broken phase while, for $\beta_{3}=1.65,\ \beta_h=0.30,$ it lies in the symmetric phase. The phase transition point, when $\beta_h=0.30,$ is estimated to be around $\beta_3\simeq 1.63.$ Thus the system is estimated to move from the broken to the symmetric phase when $\beta_3$ increases, while the remaining parameters are kept fixed. This means that big $\beta_3$ push the system towards the symmetric phase.

In fact, for phenomenological studies of dark matter, $\beta_3$ should be restricted to small values. We examine what will happen for other values of the gauge coupling $\beta_g.$ Consider, for example the two middle curves in figure \ref{C_all}, corresponding to $\beta_{3}=0.60$ and $\beta_{3}=0.70.$ If we concentrate in $\beta_g=1.0,$ we see that $\beta_{3}=0.60,\ \beta_h=0.30,$ places the system in the broken phase, while $\beta_{3}=0.70,\ \beta_h=0.30,$ places it in the symmetric phase. The estimated critical value for $\beta_{3}$ is around 0.67. Finally we consider the two curves on the left, corresponding to $\beta_{3}=0.00$ and $\beta_{3}=0.20.$ We may now easily see that the values $\beta_{g}=0.50,\ \beta_h=0.30,$ take us deep in the broken phase for $\beta_{3}=0.00$ and in the symmetric phase for $\beta_{3}=0.20,$ with an estimated critical value around $\beta_{3}=0.18.$ There follows the conclusion that critical $\beta_{3}$ values may be small, provided $\beta_g$ is small enough.

A distinctive feature of the phase diagram worth mentioning is the behaviour of the phase transition lines near $\beta_3=\beta_g.$ The coupling $\beta_g$ may not be smaller than $\beta_3.$ For instance, for $\beta_3=1.65$ (the rightmost curve), the phase transition line moves to smaller $\beta_g$ and larger $\beta_h,$ however, for large $\beta_h,$ the line becomes vertical at the value $\beta_g=1.65,$ since it may not become smaller that $\beta_3.$
In section \ref{s31} we have seen that, when $\beta_1=\beta_2=\beta_g$ and $q_{1s} = q_{2s},$ as is the case in model C, it is appropriate to describe the system through the massive $A_u^\mu$ field (which couples with the scalar field via the dimensionless charges $Q_{us}=1$) and the massless $A_v^\mu$ field, which does not couple to the scalar field, since $Q_{vs}=0.$ According to equation (\ref{BuBv}) the couplings for these fields in the symmetric phase equal: $ B_u = \beta_g-\beta_3,\ \ B_v = \beta_g+\beta_3.$ In the symmetric phase, it is evident that $\beta_g-\beta_3$ should be positive, meaning that $\beta_g$ may not be smaller than $\beta_3.$ It appears that this feature insists in the phase diagram, although we examine the broken phase and we have not done any transformation to the $A_u^\mu$ and $A_v^\mu$ fields.

The line has an abrupt change and finally it assumes its final vertical form for $\beta_h=0.38$ or larger. A rather unexpected result is that the value $\beta_h=0.38,$ for which the vertical part starts, is the same (within errors) for all $\beta_3$ values.

\begin{figure}[ht]
\begin{center}
\includegraphics[scale=0.3,angle=0]{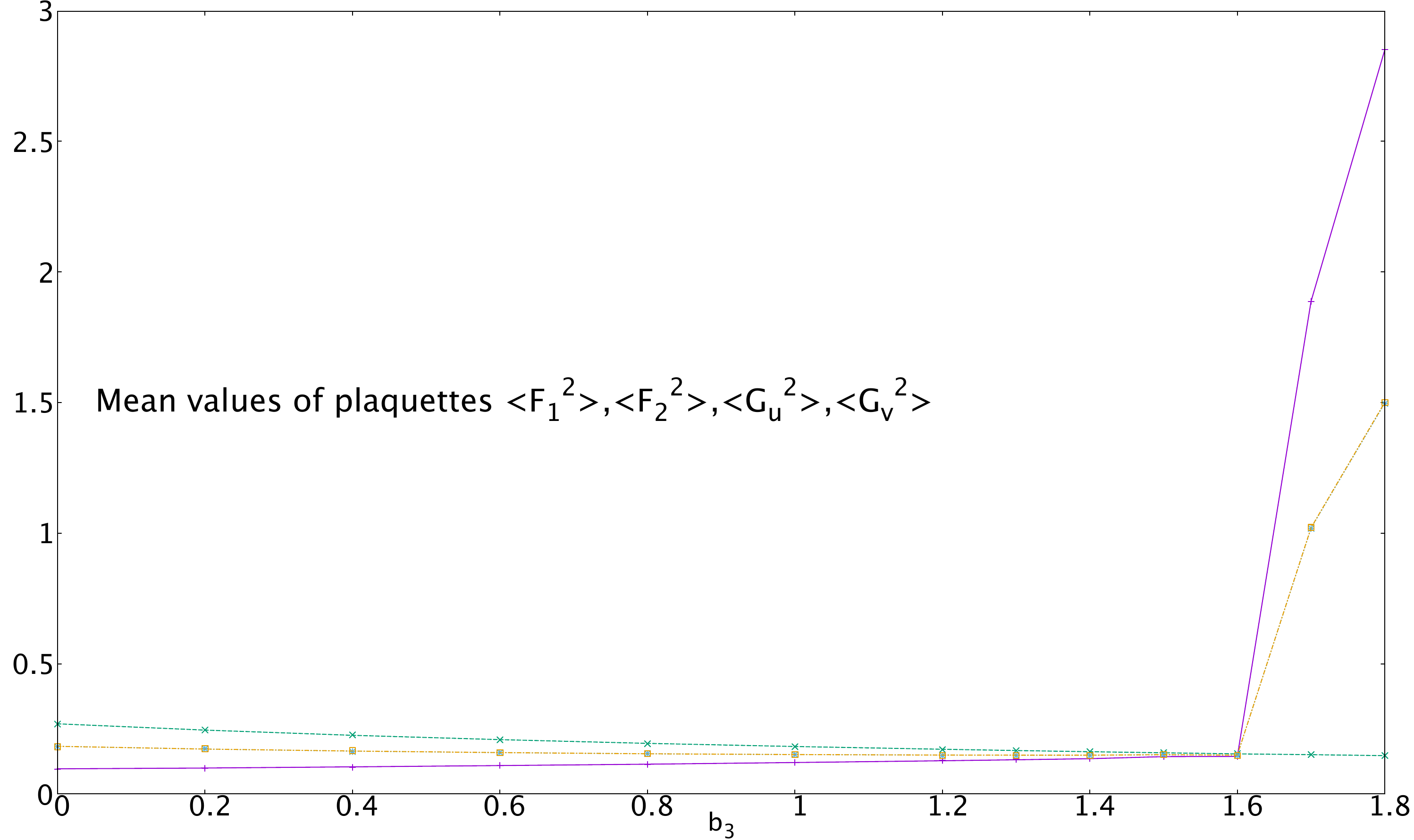}
\end{center}
\caption {Broken phase, model C: $<F^{2}_1>,$ $<F^{2}_2>$ (coincident middle curves), $<G^{2}_u>$ (curve starting at 0.11) and $<G^{2}_v>$ (curve starting at 0.27). } \label{Cf1f2gugv}
\end{figure}

We now go ahead with the study of the dependence of the quantities $<F^{2}_1>,$ $<F^{2}_2>,$ $<G^{2}_u>$ and $<G^{2}_v>$ on $\beta_3$ ($\beta_g=2.0$), which may be seen in figure \ref{Cf1f2gugv}. The variation of the various quantities is fairly mild, up to $\beta_3\simeq 1.63,$ where there is an abrupt change. %A very similar picture appears in the graph of the correlators $<F_1 F_2>$ and $<G_u G_v>$ versus $\beta_{3},$ shown in figure \ref{C_corr}.
This behaviour can be explained with the help of the phase diagram of figure \ref{C_all}. The quantities $<F^{2}_1>$ and $<F^{2}_2>$ in figure \ref{Cf1f2gugv} are equal, since their couplings to the scalar field are equal. However, we have found that the correlators $<F_1 F_2>$ cannot be neglected, so one should diagonalize the action, using the $G_u^{\mu\nu}$ and $G_v^{\mu\nu}$ fields, for which the correlator $<G_u G_v>$ vanishes. The field $G_u^{\mu\nu}$ is the one that couples to the scalar field, so that the quantity $<G^{2}_u>$ takes on rather small values. On the other hand, $G_v^{\mu\nu}$ does not couple to the scalar field, so the values of $<G^{2}_v>$  are comparatively large; they actually decrease slightly with increasing $\beta_3,$ since they are expected to be proportional to $\f{1}{\beta_g+\beta_3}.$ These explanations hold up to $\beta_3=1.63.$ At that point, as explained previously, the system passes over to the symmetric phase, so the $G_u^{\mu\nu}$ field is now proportional to $\f{1}{\beta_g-\beta_3}$ and takes very large values. This behaviour is also seen in $<F^{2}_1>$ and $<F^{2}_2>,$ since they may be expressed as functions of $<G^{2}_u>.$ On the contrary, for the $G_v^{\mu\nu}$ field nothing changes and it decreases for $\beta_3>1.63$ as well.

In table 4 the mass of $G_u^{\mu\nu}$ and the fine structure constant of $G_v^{\mu\nu},$ as well as the mass of the scalar particle for model C may be found. The fine structure constant $\alpha(G_v)$ decreases with $\beta_3,$ as expected. The mass of $G_u^{\mu\nu}$ also decreases. Actually, if one continues with the simulations for even larger $\beta_3,$ this mass will become compatible with zero and the system will pass over to the symmetric phase after $\beta_3=1.63.$ On the other hand, for small $\beta_3,$ the mass of $G_u^{\mu\nu}$ is larger than the scalar mass\cite{Mondino}; but this is reversed for $\beta_3>0.50$ and it is expected that the gauge field cannot give scalars through its decay, if $\beta_3$ grows large enough. In addition, the coupling $\alpha(G_v)$ in table 4 is approximately equal to $\alpha(G_v)$ in table 1, that is, the corresponding field in the symmetric phase. Thus the field in the symmetric phase, which is not coupled with the scalar field $(Q_{vs}=0)$ does not change essentially after symmetry breaking.

$$
    \begin{tabular} {| c | c | c | c |}
        \hline
       $\beta_3$ &  $m(G_u)$ & $\alpha(G_v)$ & $m_{scalar}$ \\
        \hline
        $0.00$   & $2.59(9) $ &  $0.0453(6)$  &  $2.11(1)$ \\
        \hline
        $0.10$   & $2.58(11)$ & $0.0435(6)$  &  $2.10(1)$  \\
        \hline
        $0.20$   & $2.52(9)$ & $0.0412(6)$   &  $2.11(1)$  \\
        \hline
        $0.30$    & $2.51(8)$ & $0.0388(4)$ &  $2.10(1)$   \\
        \hline
        $0.40$    & $2.52(10)$ & $0.0374(5)$ &  $2.09(2)$   \\
        \hline
        $0.50$    & $2.21(16)$ & $0.0354(6)$  &  $2.08(2)$  \\
        \hline
        $0.60$    & $1.60(16)$ & $0.0340(6)$ &  $2.08(2)$   \\
        \hline
        $0.80$    & $1.9(8)$ & $0.0289(6)$  &  $2.07(2)$ \\
        \hline
        $1.00$    & $0.7(3)$ & $0.0283(9)$ &  $2.06(3)$  \\
        \hline
\end{tabular}
$$
Table 4: The parameters $m(G_u)$ and $\alpha(G_v)$ in the broken phase of model C.

\section{Conclusions} \label{s5}

We have studied a model, suitable for dark matter inquiries, among others. The model is based on a product group, with $U(1)$ factors, and we have identified the resulting gauge groups in a couple of models with varying scalar field charges. The model is more general than its motivation.

(1) The tree-level mass predictions in the continuum \cite{italians} for the massive gauge fields predicted by the usual method equal the predictions of the method proposed in this work. One may see this readily by comparing (\ref{mA1}) against (\ref{amA1}) for model A, (\ref{mB1}) against (\ref{amB1}) for model B, and (\ref{mC1}) against (\ref{amC1}) for model C. That these predictions  are identical lends confidence that the approach proposed here is on the right track.

(2) The tree-level mass predictions in the continuum behave quite differently from the corresponding lattice results: they tend to diverge, when $\beta_3$ approaches $\beta_g,$ equivalently when $\kappa$ approaches one. On the contrary, their lattice counterparts approach zero in this limit, as may be seen inspecting the $m(H_1)$ column in table 3 and the the $m(G_u)$ column in table 4. The source of this strong disagreement may be traced to the fact that lattice calculations are non-perturbative, so they can, e.g., detect phase transitions, while tree-level calculations cannot. In our case, increasing $\beta_3$ to approach $\beta_g,$ moves the system towards a phase transition, which cannot be seen from the tree-level calculations.

(3) The fine structure constants displayed e.g. in tables 1, 3 and 4, are more or less the same for small enough $\beta_3$, despite the fact that they refer to different situations (for example, symmetric versus broken phases). In the symmetric phase the quantity $\alpha(G_u)$ increases, while $\alpha(G_v)$ decreases with increasing $\beta_3,$ in qualitative agreement with the tree level predictions $\alpha(G_u)\propto \f{1}{\beta_g-\beta_3},\ \alpha(G_v)\propto \f{1}{\beta_g+\beta_3}.$ In the broken phase of model B the quantity $\alpha(H_1),$ which corresponds to $\alpha(F_1),$ (this does not couple to the scalar field), is essentially constant. This is in sharp contrast with the behaviour of the fine structure constants in the symmetric phase. On the other hand, in the broken phase of model C the quantity $\alpha(G_v)$ decreases with increasing $\beta_3.$ This is qualitatively similar with its behaviour in the symmetric phase. On the contrary the field $G_u^{\mu\nu}$ yields a Yukawa behaviour.

(4) The masses in model B of the gauge field $F_2^{\mu\nu},$ which interacts with the scalar field, do not depend appreciably on $\beta_3,$ while the masses of $H_2^{\mu\nu}$ do so. These masses are equal for $\beta_3=0$ and then the mass of the gauge field $H_2^{\mu\nu}$ becomes smaller, as $\beta_3$ increases. The masses of the scalar field in model B slightly increases with $\beta_3$ and becomes bigger than the mass of the gauge field $H_2^{\mu\nu}.$ In model C the mass of the gauge field $G_u^{\mu\nu}$ is bigger than the scalar mass for $\beta_3=0,$ but it diminishes as $\beta_3$ increases and becomes smaller than the mass of the scalar field at $\beta_3> 0.50.$ The tree-level prediction for the scalar field mass is a value that does not depend on $\beta_3$ for either model.

(5) The system may move from the broken to the symmetric phase, when $\beta_3$ is changed. According to figure \ref{fig:phd170_180}, there is a phase transition from the broken to the symmetric phase in model B, as $\beta_3$ becomes bigger than 1.75. This explains the qualitatively different behaviour for $\beta_3\ge 1.75$ in figures \ref{correlB} and \ref{plaquetteb3B}. Similarly, according to figure \ref{C_all}, there is a phase transition from the broken to the symmetric phase in model C, as $\beta_3$ becomes bigger than 1.63. This explains the qualitatively different behaviour for $\beta_3\ge 1.63$ in figure \ref{Cf1f2gugv}. The critical values of $\beta_3$ move to smaller values  (for instance 0.67 or 0.18, as depicted in figure \ref{C_all}) if $\beta_g$ is chosen to be smaller, i.e. for stronger gauge coupling.
%The lattice formulation predicts a phase transition, when $\beta_3,$ corresponding to $\kappa,$ is varying.
In an attempt to relate lattice parameters with
continuum quantities we also write down the effective potential {\bf in the continuum} at one loop, where the gauge coupling $g^2$ should be replaced by $\f{g^2}{1-\kappa},$ with $g^2$ a quantity independent from $\kappa.$ This form describes the effective coupling in the presence of kinetic coupling, as explained in the paper; it is expected to provide valuable non-perturbative information.
The result is that the characteristic of two minima persists, as one increases $\kappa,$ however the separation of the two minima becomes very small. This means that the phase transition towards the symmetric phase is missed by the one-loop effective potential, but it appears that the system will possibly move to the symmetric phase near $\kappa=1,$ if higher loop corrections are included; we are working on this.

(6) We have performed simulations of the model for negative $\beta_3$ and the results for the mean values $<F_1^2>,\ <F_2^2>,\ <G_u^2>$ and $<G_v^2>$ are equal to the mean values for $|\beta_3|.$

Future work may include:\\
(a) Numerical computation for the $\beta$ functions for the various couplings via a non-perturbative calculation.
\footnote{We thank the referee for raising this issue.}
To this end one might compute the dependence of our results on the lattice size, to get an estimate of the dependence on the lattice spacing and the approach to the thermodynamic limit. This calculation would allow for a connection with physical quantities and we postpone it for a separate publication.\\
(b) Simulation of a model based on $SU(2)\times U(1)\times U(1),$ to provide the possibility to study the main physically different and realistic couplings of the dark $U(1),$ to either the hypercharge or the electromagnetic $U(1).$

\section*{Acknowledgements}

We would like to thank the administration of the ARIS Computer Center for computer time allocated to us, as well as Professor A. Tsapalis for help with the code parallelization. Thanks are also due to Dr P.Dimopoulos for illuminating discussions, as well as to professor N.E.Mavromatos for enlightening discussions and for reading the manuscript.

\end{document}